\documentclass[%
 twocolumn,
superscriptaddress,
 longbibliography,
 amsmath,amssymb,
 aps, prl,
]{revtex4-2}
\usepackage{CJK}
\usepackage{graphicx}
\usepackage{dcolumn}
\usepackage[T1]{fontenc}
\usepackage{bm}
\usepackage{gensymb}
\usepackage[dvipsnames]{xcolor}
\usepackage{amssymb,amsmath,mathptmx} 
\usepackage[colorlinks=true,citecolor=blue,linkcolor=blue,urlcolor=blue,bookmarks=false]{hyperref}
\usepackage{tikz}
\usetikzlibrary{arrows,snakes,positioning,shapes,patterns}

\def\unit#1{\mathord{\thinspace\rm #1}}

\DeclareMathAlphabet\mathbfcal{OMS}{cmsy}{b}{n}

\begin{document}

\begin{CJK*}{UTF8}{}
\title{Pseudomagnetic focusing of valley-polarized beams in strained graphene}

\author{Alina Mre\'nca-Kolasi\'nska 
}
\thanks{These two authors contributed equally.}
\affiliation{AGH University of Science and Technology, Faculty of Physics and Applied Computer Science, al. Mickiewicza 30, 30-059 Krak\'ow, Poland}

\author{Christophe De Beule}
\thanks{These two authors contributed equally.}
\affiliation{Department of Physics and Astronomy, University of Pennsylvania, Philadelphia, Pennsylvania 19104, USA}
\affiliation{Department of Physics, University of Antwerp, Groenenborgerlaan 171, 2020 Antwerp, Belgium}

\author{Jia-Tong Shi (\CJKfamily{bsmi}{施\nolinebreak佳\nolinebreak彤})}
\affiliation{Department of Physics, National Cheng Kung University, Tainan 70101, Taiwan}

\author{Aitor~Garcia-Ruiz~(\CJKfamily{bsmi}{艾\nolinebreak飛\nolinebreak宇})}
\affiliation{Department of Physics, National Cheng Kung University, Tainan 70101, Taiwan}

\author{Denis Kochan (\CJKfamily{bsmi}{可\nolinebreak汗\nolinebreak丹})}
\affiliation{Department of Physics, National Cheng Kung University, Tainan 70101, Taiwan}
\affiliation{Center for Quantum Frontiers of Research and Technology (QFort), National Cheng Kung University, Tainan 70101, Taiwan}
\affiliation{Institute of Physics, Slovak Academy of Sciences, 84511 Bratislava, Slovakia}

\author{Klaus Richter}
\affiliation{Institut f\"ur Theoretische Physik, Universit\"at Regensburg, D-93040 Regensburg, Germany}

\author{Ming-Hao Liu (\CJKfamily{bsmi}{劉明豪})}
\email{minghao.liu@phys.ncku.edu.tw}
\affiliation{Department of Physics, National Cheng Kung University, Tainan 70101, Taiwan}
\affiliation{Center for Quantum Frontiers of Research and Technology (QFort), National Cheng Kung University, Tainan 70101, Taiwan}

\date{\today}


\begin{abstract}
In addition to spin, the low-energy charge carriers in graphene have two inequivalent valley flavors $\mathbf K$ and $\mathbf K'$ related by time-reversal symmetry, which can potentially serve as carriers for additional information. One promising route of controlling the valley degree of freedom relies on strain-induced pseudomagnetic fields, which couple with opposite signs to the two valleys. In this work, we theoretically investigate quantum transport in strained graphene devices that host tailored pseudomagnetic field profiles for valley manipulation. We focus on experimentally feasible geometries and propose transverse pseudomagnetic focusing in a bent graphene ribbon as an efficient  method for generating valley-polarized currents, even in presence of disorder. Our work highlights pseudomagneto-transport in graphene as a viable tool for valleytronic applications.
\end{abstract}

\maketitle
\end{CJK*}

\section{Introduction}
The properties of two-dimensional crystals such as graphene can be significantly modified by lattice deformations. In graphene, mechanical deformations couple to the low-energy quasiparticles as scalar and gauge fields, and can be realized by applying external stress \cite{guinea_energy_2010,zhu_programmable_2015} or by placing graphene on a nanostructured substrate \cite{kiselev_optical_2011,zhang_electronic_2018,ho_hall_2021,milovanovic_strain_2019,kang_pseudo-magnetic_2021,srut_rakic_interaction_2025}. 
In particular, elastic shear strain generates pseudogauge fields with opposite signs in valley $\mathbf K$ and $\mathbf K'$ to preserve time-reversal symmetry. Consequently, certain inhomogeneous strain configurations result in pseudomagnetic fields (PMFs) with magnitudes that may exceed several hundreds of Teslas \cite{levy_strain-induced_2010,nigge_room_2019} and concomitant pseudo-Landau levels \cite{guinea_generating_2010,guinea_energy_2010,li_valley_2020,liu_analytic_2022}. Therefore, specific strain fields can be used to tailor the electronic properties of graphene through engineering PMF profiles, which is called \textit{straintronics}. For example, strain superlattices in graphene have recently been proposed as an alternative platform for topological flat bands \cite{mao_evidence_2020,milovanovic_band_2020,de_beule_network_2023,de_beule_roses_2023,de_beule_elastic_2025} and strongly-correlated phases \cite{manesco_correlations_2020,manesco_correlation-induced_2021,gao_untwisting_2023, Hsieh2026}. Moreover, because the two valleys feel opposite PMFs, they can be used to generate valley-polarized currents \cite{chaves_wave-packet_2010,nguyen_valley_2016,milovanovic_strain_2016,ortiz_graphene_2022}, thereby enabling applications in \textit{valleytronics}. Moreover, owing to their large momentum separation, intervalley scattering is strongly suppressed either by kinematic constraints, or provided that local potentials vary slowly with respect to the graphene lattice.  
Therefore, the valley degree of freedom, like spin, can be used as a medium for information processing. Notably, there exist several proposals for transport-based valley control, including valley-dependent trajectories through trigonal warping in bilayer graphene \cite{gold_coherent_2021,Ingla-Aynes_specular_2023,ingla-aynes_ballistic_2024} and valley-contrasted Berry curvature by breaking inversion symmetry to generate an intrinsic valley Hall effect in transition-metal dichalcogenides monolayers \cite{Mak_the_valley_Hall_2014}, monolayer graphene aligned to an hexagonal boron nitride substrate \cite{Gorbachev_detecting_2014}, and bilayer graphene in an electric displacement field \cite{Shimazaki_generation_2015,Sui_gate-tunable_2015}. In addition, valley valves and splitters built from topological valley Hall channels arising from band inversion in bilayer graphene \cite{Li_gate-controlled_2016,Li_a_valley_valve_2018, huang_high_temperature_2024,luna_probing_2025}. Suitably designed lattice deformations in monolayer graphene provide an alternative means of valley control. Moreover, strain-induced PMF profiles can be engineered on nanometer length scales, enabling applications that are challenging with real magnetic fields.

The prospect of graphene straintronics for next-generation devices has motivated recent experimental demonstrations of pseudomagnetotransport \cite{zhang_electronic_2018,zhang_magnetotransport_2019,zhang_gate-tunable_2022}. Inspired by this experimental progress, we study quantum transport in multi-terminal mesoscopic devices in strained graphene for realistic deformation geometries, including bent graphene ribbons and triaxially-strained hexagonal flakes \cite{guinea_energy_2010,neek-amal_electronic_2013,settnes_pseudomagnetic_2016}. Furthermore, we introduce \textit{transverse pseudomagnetic focusing} (TPMF), in analogy to transverse magnetic focusing in graphene \cite{taychatanapat_electrically_2013}. In contrast to the latter, TPMF allows one to focus a valley-polarized current through the application of strain in the absence of an externally applied magnetic field. We further show that the valley polarization of the focused current is robust against disorder, making this setup a compelling candidate for use in valleytronic applications. In addition, we also consider an S-shaped ribbon that results in a pseudomagnetic domain wall, leading to the formation of valley-chiral snake states. Both proposals display characteristic conductance oscillations that can be measured experimentally.

\section{Results and discussion}

\subsection{The model}

Throughout this paper, we model the electronic properties of graphene with the tight-binding Hamiltonian describing electrons hopping between nearest-neighbor carbon $p_z$ orbitals with amplitude $t_{mn}$ in the presence of an onsite potential $U_n$ due to applied gate voltages,
\begin{equation} \label{eq:Htb}
    \hat H = -\sum\limits_{\left\langle {m,n} \right\rangle } t_{mn} \hat c_m^\dagger \hat c_n + \sum\limits_n U_n \hat c_n^\dagger \hat c_n,
\end{equation}
where $\hat c_n^\dag$ ($\hat c_n$) are operators that create (destroy) an electron on the $n$-th site. In the presence of in-plane deformations, the original position of the $n$-th site $\mathbf{r}_n = (x_n,y_n)$ becomes $\mathbf{R}_n = \mathbf{r}_n + \mathbf{u}(\mathbf r_n)$ where $\mathbf u(x,y)$ is the in-plane (acoustic) displacement field \cite{vozmediano_gauge_2010}. These lattice deformations couple to the charge carriers by modifying the hopping amplitudes. Working in the central-force approximation, we use a phenomenological hopping function given by \cite{pereira_tight-binding_2009}
\begin{equation}
    t_{mn} = t_0 \exp\left[ -\beta \left( \frac{|\mathbf R_m - \mathbf R_n|}{a_0} - 1 \right) \right],
    \label{eq tij}
\end{equation}
where $t_0 = 3\unit{eV}$, $a_0 = (4\sqrt{3})^{-1}\unit{nm}$ is the nearest-neighbor distance of pristine graphene, and we take $\beta = 3.37$.

To make the transport simulations for mesoscopic strained graphene devices computationally feasible, we use the scaling method for unstrained graphene from Ref.\ \cite{liu_scalable_2015} and its generalization to account for strain \cite{liu_2026_scalable}. In this way, we can reduce the Hamiltonian matrix size, while leaving the long-wavelength physics near the charge neutrality point, that we are interested in, unchanged. Specifically, the generalized scaling transformation consists of properly scaling the lattice constant, hopping amplitude, and the displacement field, resulting in the following scaling laws:
\begin{align} \label{eq:scaling}
     a_0 \rightarrow s a_0\ , && t_0 \rightarrow t_0/s\ , && \mathbf u \rightarrow s \mathbf u\ ,
\end{align}
which leave the low-energy physics invariant as long as the scaled strain remains in the elastic regime \cite{liu_2026_scalable}. Here, we use continuum elasticity to model the strain. Although this only accounts for acoustic displacements, the optical component describing relative motion between different sublattices is effectively included by a reduction factor that renormalizes the electron Gr\"uneisen parameter $\beta$ \cite{woods_electron-phonon_2000,suzuura_phonons_2002,de_beule_elastic_2025}. Finally, in our quantum transport calculations, the main object of interest is the conductance from lead $l$ to lead $k$, which is obtained from the Landauer formula $G_{kl}= 2(e^2/h) T_{kl}$ where the transmission function $T_{kl}$ is calculated numerically with the wave-function-matching method  within the linear response regime in the limit of zero temperature \cite{zwierzycki_calculating_2008, kolasinski_interference_2016}. 

\subsection{Transverse pseudomagnetic focusing}

\begin{figure}[t!]
    \centering
    \includegraphics[width=\columnwidth]{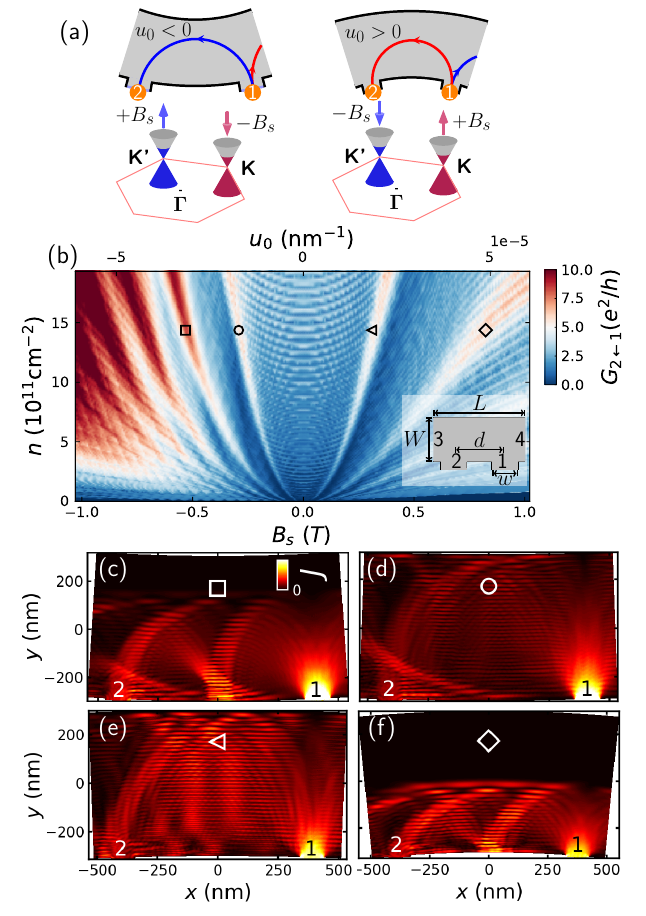}
    \caption{\textbf{Transverse pseudomagnetic focusing.} (a) Schematic of the proposed device using a bent graphene zigzag ribbon. We also show the Dirac cones at $\mathbf{K}$ and $\mathbf{K'}$ together with the PMF sign and electron trajectories in red and blue, respectively. (b) Injector to collector conductance as a function of bending strength $u_0$ and carrier density $n$ set by a global back gate, using scaling $s=2$. The lower horizontal axis gives the corresponding PMF. The inset shows the labeling of the leads in the device. The smaller oscillations, most prominent between the first two focusing peaks, are Fabry-P\'erot-like oscillations due to transverse confinement in the ribbon. (c)--(f) Current density maps at selected conductance peaks, marked by the symbols in (b).}
    \label{fig1}
\end{figure}

The main focus of this work is a multi-terminal device made from a bent graphene nanoribbon, as shown in Fig.\ \ref{fig1}(a). The corresponding displacement fields are  given by \cite{guinea_generating_2010}
\begin{equation}
    \begin{aligned}
    u_x &= u_0 \left(2xy + \frac{f_0}{f_1}x \right), \\
    u_y &= u_0 \left[ \frac{\nu}{\nu - 2} \left( y^2 + \frac{f_0}{f_1}y \right) - x^2 \right],
    \end{aligned}
    \label{eq ur uy for bent graphene}
\end{equation}
where $f_0$ and $f_1$ are constants describing forces acting on the right and left boundaries, the Poisson ratio of graphene $\nu = 0.15$, and $u_0$ with units of inverse length controls the bending strength. The corresponding strain profile results in a uniform PMF $B_s = \tfrac{2\hbar\beta}{ea_0} \tfrac{u_0}{2 - \nu} \cos(3\theta)$, independent of $f_0/f_1$ which only contributes a pure gauge \cite{guinea_generating_2010,chaves_wave-packet_2010,Note1}. Here, $\theta$ is the angle between the zigzag direction and the $x$ axis. In the remainder of this work, we consider $\theta = 0$. This setup lends itself naturally to a pseudomagnetic focusing geometry, in analogy to transverse magnetic focusing (TMF) \cite{taychatanapat_electrically_2013}. In a typical TMF geometry, two narrow probes are located on the same edge of a planar device, acting as a current injector and collector, while other leads serve as drains for the current that does not flow into the collector. When an external magnetic field perpendicular to the plane is applied in the absence of strain, semiclassically, the injected charge carriers can reach the collector only for commensurate trajectories, namely if an integer multiple of the diameter of the cyclotron orbit fits in between the injector and the collector \cite{beenakker_mode_1988,van-Houten_coherent_1989}.

For the transport calculation, we take a system of width $W=600$~nm and length $L=1029$~nm. The injector and collector leads have widths $w = 100$~nm and are spaced by $d = 800$~nm, and we use a scaling factor $s=2$. The system contains four leads in total, as shown in the inset of Fig.\ \ref{fig1}(b). In our TPMF setup, the magnitude of the PMF, denoted by $B_s$, is proportional to the bending strength which is quantified by the parameter $u_0$ in Eq.\ \eqref{eq ur uy for bent graphene}. See Supporting Information (SI) for details \cite{Note1}. The sign of $B_s$ corresponds to bending inwards or outwards, see Fig.\ \ref{fig1}(a), resulting in the pseudo-cyclotron trajectories schematically depicted in the figure. For trajectories commensurate with the lead spacing, a peak in the conductance is expected. Unlike a real magnetic field, where carriers of a given charge are focused only for a given direction of the magnetic field \cite{taychatanapat_electrically_2013}, here for a given PMF with $u_0>0$ ($u_0<0$) only $\mathbf{K}$ ($\mathbf{K'}$) electrons are focused. Indeed, in Fig.\ \ref{fig1}(b) we observe peaks in the conductance between the injector (lead $1$) and the collector (lead $2$) for both signs of $B_s$. These focusing peaks are not symmetric around $B_s=0$ due to the change in lead distance for bending inwards and outwards, see Fig.\ \ref{fig1}(a). For $u_0>0$, the probe spacing is reduced and higher $B_s$ is needed for commensurate cyclotron orbits. Conversely, for $u_0<0$, the probes are further apart, requiring lower $B_s$ and leading to more densely spaced focusing peaks. The smaller oscillations with carrier density that appear most prominently between the first two focusing peaks are Fabry-P\'erot-like oscillations due to transverse confinement in the ribbon. They disperse quadratically with the PMF because the phase $k_F W$ is modified by $\sim -W^3 / (k_F \ell_s^4)$, where $\ell_s=\sqrt{\frac{\hbar}{e|B_s|}}$ is the pseudo-magnetic length.
To further illustrate the TPMF, we show the magnitude of the total current density summed over all scattering states in Fig.~\ref{fig1}(c)--(f) corresponding to different conductance peaks marked by the symbols in Fig.\ \ref{fig1}(b).
The corresponding displacement profile can be approximated by an arc-like deformation: $u_x = xy/R$, $u_y = -x^2/2R$, where $R$ is the radius of the inner circle defining the lower edge; then we can estimate $R\approx 1/2u_0$, following \cite{guinea_generating_2010}. For the maximum of $u_0=6\times 10^{-5}$ nm$^{-1}$, this gives an estimation of $R\approx 8.3\ \mu$m
\begin{figure}[t!]
    \centering
    \includegraphics[width=\columnwidth]{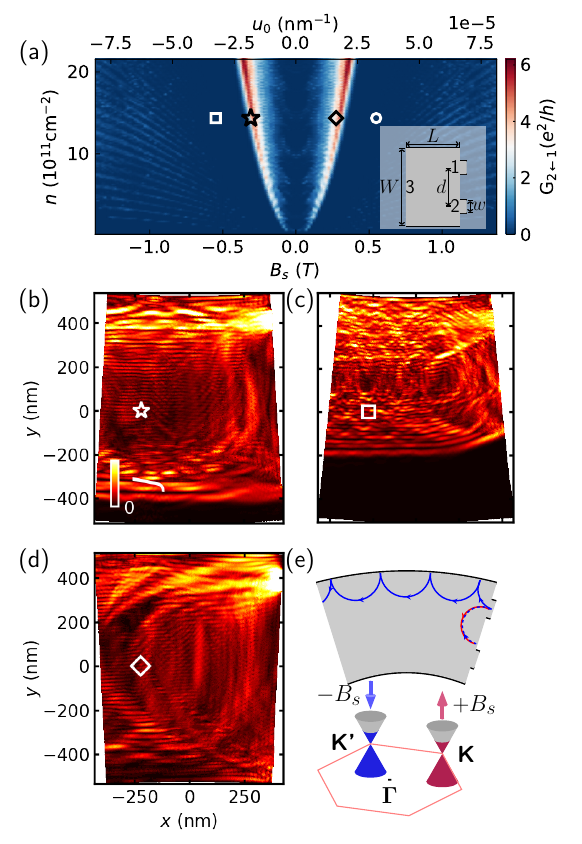}
    \caption{\textbf{Intervalley scattering at the armchair edge.} (a) Injector to collector conductance as a function of $u_0$ and carrier density. The lower $x$-axis is the strain converted to pseudomagnetic field. (b)--(d) Current density maps at $n$ and $B_s$ values marked by respective symbols. (e) Illustration of backscattering due to intervalley scattering at the armchair edge. The red (blue) trajectory represents $\mathbf{K}$ ($\mathbf{K}'$) valley electrons, and the dashed blue is the reversed trajectory after scattering from $\mathbf{K}$ to $\mathbf{K}'$.}
    \label{fig2}
\end{figure}

Note that we have considered a ribbon for which the injector and collector leads lie along a zigzag edge, which precludes intervalley scattering, as in this case the two valleys correspond to different transverse momenta. To address this, we have performed a similar calculation for an armchair edge where intervalley scattering is kinematically allowed. 
For this, we consider a similar setup but with the injector and collector placed on the vertical, armchair side [Fig.\ \ref{fig2}]. This three-terminal system is of a different size than the four-terminal setup shown in Fig.\ \ref{fig1}(a). Specifically, we now take $W=400$~nm, $L=520$~nm, side-lead width $w=50$~nm and spacing $d=400$~nm, and scaling $s=4$. 
Fig.\ \ref{fig2}(a) shows the transmission from the injector to the collector as a function of the PMF and carrier density. In contrast to the previous case, where the injector and collector where placed on the zigzag edge, here only the first valley-polarized focusing peaks are present.
Examples of the current density for the direct focusing between injector and collector are shown in Fig.\ \ref{fig2}(b) and Fig.\ \ref{fig2}(d), marked by the $\star$ and $\diamond$ symbols in Fig.\ \ref{fig2}(a), respectively. When $B_s$ is increased beyond the first focusing peak, the current hits the edge between the side leads [Fig.\ \ref{fig2}(c)] and the orbits cease to propagate further along the vertical edge. This behavior results from intervalley scattering. At the armchair edge, states from $\mathbf K$ are retro-reflected to $\mathbf K'$ and vice versa, and no higher-order focusing peaks appear in the conductance. This is illustrated by the red (blue dashed) trajectory for $\mathbf K$ ($\mathbf K'$) in Fig.\ \ref{fig2}(e).

The zigzag and armchair edge terminations are two limiting cases. 
The termination of real samples consists in general of both zigzag-like and armchair-like segments, leading to some intervalley scattering between forward- and backward-moving skipping orbits, and a suppression of the higher TPMF peaks in the conductance. Also, achieving a uniform PMF requires bending a zigzag ribbon as shown in Fig.\ \ref{fig1}(a). Bending along a different direction reduces the PMF, requiring a stronger deformation to reach the same $B_s$ values that satisfy the focusing conditions. And bending an armchair ribbon produces no PMF \cite{guinea_generating_2010}. Apart from this extreme case, the first and higher (if present) focusing peaks remain valley polarized. Hence, this results in a highly valley-polarized current reaching the collector in our proposed TPMF setup.

\begin{figure}[ht!]
    \centering
    \includegraphics[width=\columnwidth]{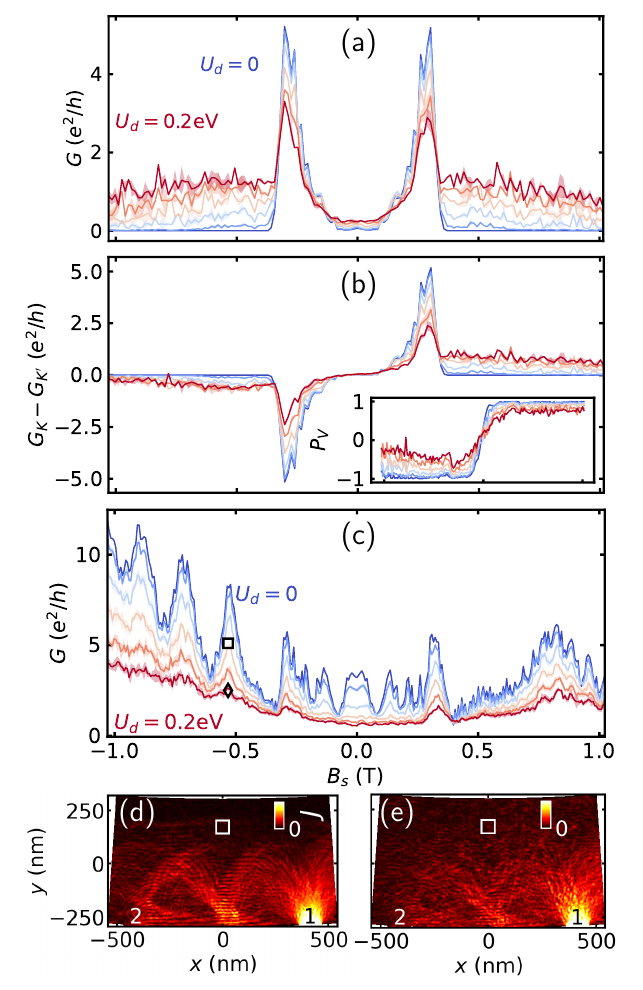}
    \caption{\textbf{Effect of disorder on TPMF.} (a) Injector to collector conductance as a function of $B_s$ and disorder strength in the system with side leads along the armchair edge [Fig.\ \ref{fig2}(e)]. Here $U_d$ is incremented from 0 to 0.2 eV in steps of 0.04 eV. (b) $G_{\mathrm{K}}-G_{\mathrm{K'}}$ at the same $U_d$ as in (a). Inset shows the corresponding $P_V$. (c) Same as (a) in the system with side leads along the zigzag edge [Fig.\ \ref{fig1}(a)] where $U_d$ ranges from 0 to 0.2 eV by 0.04 eV. The results in (a)--(c) are averaged over four random disorder configurations. The shaded regions show the interval $\pm \sigma$, where $\sigma$ is the standard deviation. (d)--(e) Current density maps at (d) $U_d=0.12$eV and (e) $U_d=0.2$eV for the $n$ and $B_s$ values marked in (c) as {\scriptsize $\square$} and $\lozenge$, respectively.}
    \label{fig3}
\end{figure}

Up to now, we considered ballistic samples where intervalley scattering only occurs at an edge. However, in general any perturbation on the atomic scale may cause valley mixing. Surprisingly, we find that the TPMF peaks are quite robust against considerably strong disorder, as shown in Fig.\ \ref{fig3}. Here, short-range scatterers are accounted for in the Hamiltonian (\ref{eq:Htb}) by uniformly distributed random onsite energies of magnitude $|U_n|<U_d$ with $s=2$, and the transport result is averaged over four disorder configurations. The shaded regions showing the confidence interval (within $\pm \sigma$, where $\sigma$ is the standard deviation) for each line cut. In Fig.\ \ref{fig3}(a), we show the conductance at density $n=14.4\times 10^{11}$~cm$^{-2}$ as a function of $B_s$ for the setup presented in Fig.\ \ref{fig2} for several values of the disorder strength $U_d$. The increasing disorder amplitude produces a background conductance since the scattered current can now also reach the collector in the absence of commensuration. Nevertheless, the focusing peaks are robust against the considered short-range disorder. 
For the armchair edge, the side leads extend in the zigzag direction, so it is possible to calculate the contributions from modes belonging to $\mathbf{K}$ and $\mathbf{K'}$, denoted as $G_{\mathrm{K}}$ and $G_{\mathrm{K'}}$, respectively. In Fig.\ \ref{fig3}(b) we plot $G_{\mathrm{K}}-G_{\mathrm{K'}}$ for the same range as in Fig.\ \ref{fig3}(a), and in the inset of Fig.\ \ref{fig3}(b)--the corresponding $P_V=(G_{\mathrm{K}}-G_{\mathrm{K'}})/G$. These results confirm the dominant contribution of the $\mathbf{K}$ ($\mathbf{K'}$) valley for $B_s>0$ ($B_s<0$), in particular near the focusing peaks. In the SI \cite{Note1} we also analyze the effect of edge roughness that causes partial valley mixing and restores the higher focusing peaks in the armchair geometry. In Fig.\ \ref{fig3}(c) we present similar calculations for the conductance in the presence of disorder for the system shown in Fig.\ \ref{fig1}(a). 
With increasing $U_d$, the overall conductance decreases, but the focusing peaks stand out even at elevated disorder, and only disappear at $U_d\approx0.2$~eV. Figures \ref{fig3}(d)--(e) show maps of the current density for the peaks marked by {\scriptsize $\square$} and $\lozenge$, respectively, in Fig.\ \ref{fig3}(c). Note that the additional oscillations between the first focusing peaks for $|B_s|\lesssim 0.25$~T due to transverse quantization, decay much faster with increasing $U_d$. We attribute the robustness to disorder of TPMF, to the valley-contrasting cyclotron motion. Semiclassicaly, when an electron in a pseudomagnetic skipping orbit scatters between valleys, the propagation direction reverses, so it does not contribute to the collected current. Thus, the current reaching the probe is reduced but remains valley polarized.

So far, we have considered systems that realize TPMF for a uniform PMF. We have also investigated the influence of small deviations from uniformity, which can occur for other strain profiles. As an example, we consider triaxially-strained graphene (see SI for details \cite{Note1}). Similarly as before, the local current densities maps reveal clear pseudomagnetic focusing of carriers, separated into two beams belonging to opposite valleys, which generate a valley-polarized beam that reaches the collector. Furthermore, at lower densities, the system is in the pseudo-Hall regime, and the current densities show two valley-chiral branches of edge states. Transport in the presence of a pseudomagnetic field is then compared with results obtained for a real magnetic field and without strain, where the edge current propagates in the same direction for both valleys. We conclude that as long as the local pseudomagnetic orbit does not change significantly along the focusing trajectory, deviations from a uniform PMF do not compromise the valley-polarized focusing current.

\subsection{Pseudomagnetic snake states}

\begin{figure}[t!]
    \centering
    \includegraphics[width=\columnwidth]{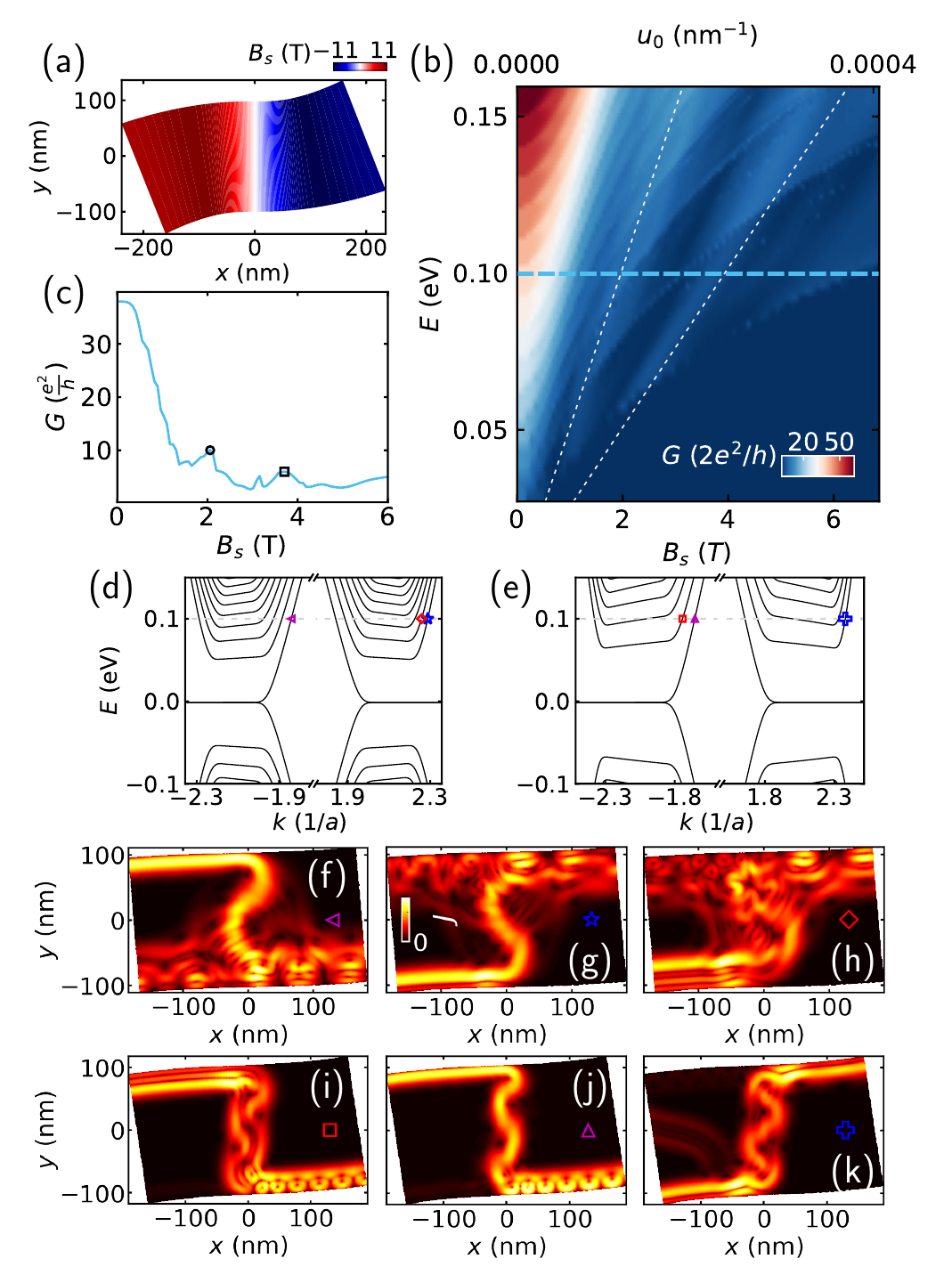}
    \caption{\textbf{Pseudomagnetic snake states.} (a) Pseudomagnetic field in an S-shaped bent zigzag graphene ribbon with scaling $s=2$. (b) Two-terminal conductance $G$ for the bent ribbon shown in (a) as a function of $u_0$ or $B_s$ and Fermi energy. The thin white dashed lines show the commensurability condition $2nR_c = W$. (c) Conductance line cut along the dashed blue line in (b) as a function of the PMF magnitude in the left and right regions. (d) Lead dispersion for $u_0$ corresponding to the first conductance peak marked by $\circ$ in (c), and (e) for the second peak marked by {\scriptsize $\square$}. (f)--(h) Current density maps at the first $G$ peak of modes marked in (d) by the respective symbols. (i)--(k) Current density maps at the second $G$ peak of modes marked in (e).}
    \label{fig4}
\end{figure}

In addition to TPMF, which requires a nearly uniform PMF, we also consider devices with spatially-varying PMF profiles. As an example, we perform two-terminal transport simulations for S-bent ribbons. This setup gives rise to a pseudomagnetic domain wall where the PMF changes sign along the transport direction. Pseudomagnetic snake states localized at the interface \cite{liu_snake_2015, rickhaus_snake_2015, taychatanapat_conductance_2015} then give rise to characteristic conductance oscillations. To achieve a highly non-uniform PMF in practice, we consider a spatially-varying bending strength by letting $u_0 \rightarrow -u_0 \tanh[ x / (2d_*) ]$ [see Eq.\ (\ref{eq ur uy for bent graphene})]. This smoothly interpolates between bending the ribbon inwards and outwards, resulting in the deformation and PMF shown in Fig.\ \ref{fig4}(a). Hence, the PMF changes sign at $x=0$. In analogy to a bipolar junction in graphene in a uniform real magnetic field \cite{liu_snake_2015, rickhaus_snake_2015, taychatanapat_conductance_2015}, the reversal of the PMF for a uniform carrier density gives rise to a snake state bound at the interface. The difference is that the snake states arise here due to the alternating PMF, rather than the sign change of the quasiparticle charge. 

In Fig.\ \ref{fig4}(b) we show the two-terminal conductance map $G(u_0,E)$ as a function of strain $u_0$ and Fermi energy for an S-shaped bent zigzag graphene ribbon of width $W\approx 206$~nm and length $L\approx 350$~nm, scaling $s=2$, and domain wall width $d_*=12$~nm. It displays oscillations as a function of both $E$ and $u_0$. The latter are shown more clearly in the line cut in Fig.\ \ref{fig4}(c), which is taken along the dashed light-blue line in Fig.\ \ref{fig4}(b). The maximums of $G$ fall close to the dashed while lines in Fig.\ \ref{fig4}(b), marking the commensurability condition $2 n R_c = W$ with $n=2,4$, where the cyclotron radius $R_c=E/e v_F B_s$ fits $2n$ times in the ribbon. Analysis of the mode-resolved current density at the first [marked by $\circ$ in Fig.\ \ref{fig4}(c)] and the second local maximum of $G$ [marked by {\scriptsize $\square$} in Fig.\ \ref{fig4}(c)] reveals snake orbits as the underlying cause of the oscillations, as we describe below. Figures \ref{fig4}(d) and \ref{fig4}(e) show the energy bands of the left lead taken for the values of $u_0$ at the first and second conductance peaks, respectively. Here, the current densities of the propagating modes marked in Figs.\ \ref{fig4}(d) and (e) are plotted in Figs.\ \ref{fig4}(f)--(k). The current incoming from the left lead ($x<0$) propagates along the lower or upper edge for the mode belonging to the $\mathbf{K}$ [Figs.\ \ref{fig4}(g), (h) and (k)] or $\mathbf{K}'$ valley [Figs.\ \ref{fig4}(f), (i), and (j)], respectively. As the current reaches the interface at $x=0$ where the PMF changes sign, it tunnels into the pseudomagnetic snake state, which is seen as alternating half-circles along the interface. When it reaches the opposite edge of the ribbon, it splits into two parts going into the left or right lead. In Figs.\ \ref{fig4}(f)--(h), the current trajectories feature two half-circles at the interface before the current flows into the right lead. In contrast, in Figs.\ \ref{fig4}(i)--(k) the number of half-circles is doubled because the magnitude of the PMF in both regions is almost doubled. For both cases, the pseudomagnetic snake states form a commensurate trajectory across the ribbon, which is the source of the conductance peaks in Figs.\ \ref{fig4}(b) and (c).

We note that our setup is similar to a sign flip of a real magnetic field, which was realized in a two-dimensional GaAs/AlGaAs electron gas \cite{ye_electrons_1995}. However, in our case, the pseudomagnetic snake states propagate in opposite directions along the interface for valley $\mathbf K$ and $\mathbf K'$. Moreover, because they are realized purely by strain, the PMF domain wall is much sharper \cite{cavalcante_all-strain_2016, Jones_Quantized_2017}, and thus the snake states are more localized at the interface. This exemplifies the opportunities that strain engineering offers for the development of electronic devices.

\subsection{Conclusions}

In summary, we demonstrated how valley-polarized currents can be achieved by transverse pseudomagnetic focusing in a multi-terminal device made from a bent graphene ribbon. This setup is based on the valley-chiral skipping orbits which results in a spatial separation of valley-polarized beams. Hence, it is robust against moderately strong disorder. Although preferentially, the bending should be in the zigzag orientation which coincides with the $x$ direction for our choice of coordinates to achieve a uniform pseudomagnetic field, as long as the local pseudomagnetic cyclotron radius does not vary significantly on the scale of the spacing between the injector and collector, the functionality of the system is not compromised by small deviations from a uniform pseudomagnetic field. This allows for more realistic deformation profiles that deviate from the fine-tuned displacement field (\ref{eq ur uy for bent graphene}) for circular-arc bending \cite{guinea_generating_2010}. 
Finally, we identified a fingerprint of pseudomagnetic snake states, localized at a domain wall where the pseudomagnetic field changes sign, in conductance oscillations for quantum transport across an S-shaped zigzag graphene ribbon. 

Recent progress in engineering the strain profile in graphene systems \cite{kapfer_programming_2023,pantaleon_designing_2024,ma_giant_2025, Zhou2025} complemented by extensive theoretical predictions \cite{DeBeule2026} may facilitate future experimental demonstration of our findings. Our work opens new avenues for applications of valley-based information processing in graphene straintronic devices.

%
%


\begin{acknowledgments}
A.M.-K. acknowledges the support of the National Science Centre, Poland, under research project no 2024/55/D/ST3/00538, and by Polish high-performance computing infrastructure PLGrid (HPC Center: ACK Cyfronet AGH) for providing computer facilities and support within computational grant no.\ PLG/2024/017407. C.D.B.\ was supported by the U.S.\ Department of Energy under Grant No.\ DE-FG02-84ER45118. J.-T.S, A.G.-R., and M.-H.L.\ acknowledge the National Center for High-performance Computing (NCHC) for providing computational and storage resources, and the National Science and Technology Council (NSTC) of Taiwan, Grant No.\ 112-2112-M-006-019-MY3, 113-2918-I-006-004, 114-2112-M-006-029-MY3, 114-2811-M-006-038, for financial support. D.K. acknowledges Taiwan National Science and Technology Council Grant No.~114-2112-M-006-034-MY3, Slovak Academy of Sciences Project IMPULZ-2021-26 SUPERSPIN, and Slovak Research and Development Agency Contract No.~APVV-24-0134.
K.R.\ acknowledges funding through the Deutsche Forschungsgemeinschaft (DFG, German Research Foundation) within Project-ID 314695032 -- SFB 1277. It furthermore supported, together with the Alumni Program of the Alexander von Humboldt Foundation, M.-H.L.'s sabbatical stay at the University of Regensburg, where part of this work was done.
\end{acknowledgments}



\noindent The simulation data underlying this study are openly available in the Rodbuk repository

\noindent  https://doi.org/10.58032/AGH/Y6RAPK



\bibliography{references_}

@misc{Hsieh2026,
      title={Emergent frustrated magnetism in strain-patterned graphene}, 
      author={Yu-Chiang Hsieh and Wen-Han Kao and Christophe De Beule and Sheng-Zhu Ho and Ru-Long Gou and Bo-Nian Chen and Kuan-Yu Chou and Kuo-En Chang and Chin-Chia Chang and Ying-Mei Yang and Ching-Hua Kao and Hao-Chien Chiang and Jyun-Lin Chen and Sheng-Chin Ho and Kenji Watanabe and Takashi Taniguchi and Ming-Hao Liu and Ching-Hao Chang and Yi-Chun Chen and Ying-Jer Kao and Tse-Ming Chen},
      year={2026},
      eprint={2609.31581},
      archivePrefix={arXiv},
      primaryClass={cond-mat.mes-hall},
      url={https://arxiv.org/abs/2609.31581}, 
}

@Article{Mrenca2023_probing,
author={Mre{\'{n}}ca-Kolasi{\'{n}}ska, Alina
and Chen, Szu-Chao
and Liu, Ming-Hao},
title={Probing miniband structure and Hofstadter butterfly in gated graphene superlattices via magnetotransport},
journal={npj 2D Materials and Applications},
year={2023},
month={Sep},
day={09},
volume={7},
number={1},
pages={64},
issn={2397-7132},
doi={10.1038/s41699-023-00426-9},
url={https://doi.org/10.1038/s41699-023-00426-9}
}

@article{DeBeule2026,
  title = {Gate-tunable resonances and one-dimensional channel in a graphene nanoslide},
  author = {De Beule, Christophe and Liu, Ming-Hao and Partoens, Bart and Covaci, Lucian},
  journal = {Phys. Rev. B},
  volume = {113},
  issue = {20},
  pages = {L201403},
  numpages = {6},
  year = {2026},
  month = {May},
  publisher = {American Physical Society},
  doi = {10.1103/g6fn-xnf3},
  url = {https://link.aps.org/doi/10.1103/g6fn-xnf3}
}

@Article{Shimazaki_generation_2015,
    author={Shimazaki, Y.
    and Yamamoto, M.
    and Borzenets, I. V.
    and Watanabe, K.
    and Taniguchi, T.
    and Tarucha, S.},
    title={Generation and detection of pure valley current by electrically induced Berry curvature in bilayer graphene},
    journal={Nature Physics},
    year={2015},
    month={Dec},
    day={01},
    volume={11},
    number={12},
    pages={1032-1036},
    issn={1745-2481},
    doi={10.1038/nphys3551},
    url={https://doi.org/10.1038/nphys3551}
}

@Article{luna_probing_2025,
	title={{Probing valley phenomena with gate-defined valley splitters}},
	author={Juan Daniel Torres Luna and Kostas Vilkelis and Antonio L. R. Manesco},
	journal={SciPost Phys.},
	volume={18},
	pages={062},
	year={2025},
	publisher={SciPost},
	doi={10.21468/SciPostPhys.18.2.062},
	url={https://scipost.org/10.21468/SciPostPhys.18.2.062},
}

@article{
Gorbachev_detecting_2014,
author = {R. V. Gorbachev  and J. C. W. Song  and G. L. Yu  and A. V. Kretinin  and F. Withers  and Y. Cao  and A. Mishchenko  and I. V. Grigorieva  and K. S. Novoselov  and L. S. Levitov  and A. K. Geim },
title = {Detecting topological currents in graphene superlattices},
journal = {Science},
volume = {346},
number = {6208},
pages = {448-451},
year = {2014},
doi = {10.1126/science.1254966},
URL = {https://www.science.org/doi/abs/10.1126/science.1254966},
eprint = {https://www.science.org/doi/pdf/10.1126/science.1254966}}

@article{
Mak_the_valley_Hall_2014,
author = {K. F. Mak  and K. L. McGill  and J. Park  and P. L. McEuen },
title = {The valley Hall effect in MoS2 transistors},
journal = {Science},
volume = {344},
number = {6191},
pages = {1489-1492},
year = {2014},
doi = {10.1126/science.1250140},
URL = {https://www.science.org/doi/abs/10.1126/science.1250140},
eprint = {https://www.science.org/doi/pdf/10.1126/science.1250140}}

@Article{Li_gate-controlled_2016,
    author={Li, Jing
    and Wang, Ke
    and McFaul, Kenton J.
    and Zern, Zachary
    and Ren, Yafei
    and Watanabe, Kenji
    and Taniguchi, Takashi
    and Qiao, Zhenhua
    and Zhu, Jun},
    title={Gate-controlled topological conducting channels in bilayer graphene},
    journal={Nature Nanotechnology},
    year={2016},
    month={Dec},
    day={01},
    volume={11},
    number={12},
    pages={1060-1065},
    issn={1748-3395},
    doi={10.1038/nnano.2016.158},
    url={https://doi.org/10.1038/nnano.2016.158}
}

@Article{Sui_gate-tunable_2015,
    author={Sui, Mengqiao
    and Chen, Guorui
    and Ma, Liguo
    and Shan, Wen-Yu
    and Tian, Dai
    and Watanabe, Kenji
    and Taniguchi, Takashi
    and Jin, Xiaofeng
    and Yao, Wang
    and Xiao, Di
    and Zhang, Yuanbo},
    title={Gate-tunable topological valley transport in bilayer graphene},
    journal={Nature Physics},
    year={2015},
    month={Dec},
    day={01},
    volume={11},
    number={12},
    pages={1027-1031},
    issn={1745-2481},
    doi={10.1038/nphys3485},
    url={https://doi.org/10.1038/nphys3485}
}

@article{huang_high_temperature_2024,
author = {Ke Huang  and Hailong Fu  and Kenji Watanabe  and Takashi Taniguchi  and Jun Zhu },
title = {High-temperature quantum valley Hall effect with quantized resistance and a topological switch},
journal = {Science},
volume = {385},
number = {6709},
pages = {657-661},
year = {2024},
doi = {10.1126/science.adj3742},
URL = {https://www.science.org/doi/abs/10.1126/science.adj3742},
eprint = {https://www.science.org/doi/pdf/10.1126/science.adj3742}}

@article{Li_a_valley_valve_2018,
author = {Jing Li  and Rui-Xing Zhang  and Zhenxi Yin  and Jianxiao Zhang  and Kenji Watanabe  and Takashi Taniguchi  and Chaoxing Liu  and Jun Zhu },
title = {A valley valve and electron beam splitter},
journal = {Science},
volume = {362},
number = {6419},
pages = {1149-1152},
year = {2018},
doi = {10.1126/science.aao5989},
URL = {https://www.science.org/doi/abs/10.1126/science.aao5989},
eprint = {https://www.science.org/doi/pdf/10.1126/science.aao5989}}

@article{gold_coherent_2021,
  title = {Coherent Jetting from a Gate-Defined Channel in Bilayer Graphene},
  author = {Gold, Carolin and Knothe, Angelika and Kurzmann, Annika and Garcia-Ruiz, Aitor and Watanabe, Kenji and Taniguchi, Takashi and Fal'ko, Vladimir and Ensslin, Klaus and Ihn, Thomas},
  journal = {Phys. Rev. Lett.},
  volume = {127},
  issue = {4},
  pages = {046801},
  numpages = {6},
  year = {2021},
  month = {Jul},
  publisher = {American Physical Society},
  doi = {10.1103/PhysRevLett.127.046801},
  url = {https://link.aps.org/doi/10.1103/PhysRevLett.127.046801}
}

@article{ingla-aynes_ballistic_2024,
  title = {Ballistic Electron Source with Magnetically Controlled Valley Polarization in Bilayer Graphene},
  author = {Ingla-Ayn\'es, Josep and Manesco, Antonio L. R. and Ghiasi, Talieh S. and Watanabe, Kenji and Taniguchi, Takashi and van der Zant, Herre S. J.},
  journal = {Phys. Rev. Lett.},
  volume = {133},
  issue = {15},
  pages = {156301},
  numpages = {7},
  year = {2024},
  month = {Oct},
  publisher = {American Physical Society},
  doi = {10.1103/PhysRevLett.133.156301},
  url = {https://link.aps.org/doi/10.1103/PhysRevLett.133.156301}
}

@Article{Ingla-Aynes_specular_2023,
    author={Ingla-Ayn{\'e}s, Josep
    and Manesco, Antonio L. R.
    and Ghiasi, Talieh S.
    and Volosheniuk, Serhii
    and Watanabe, Kenji
    and Taniguchi, Takashi
    and van der Zant, Herre S. J.},
    title={Specular Electron Focusing between Gate-Defined Quantum Point Contacts in Bilayer Graphene},
    journal={Nano Letters},
    year={2023},
    month={Jun},
    day={28},
    publisher={American Chemical Society},
    volume={23},
    number={12},
    pages={5453-5459},
    issn={1530-6984},
    doi={10.1021/acs.nanolett.3c00499},
    url={https://doi.org/10.1021/acs.nanolett.3c00499}
}

@article{kiselev_optical_2011,
	title = {Optical and photogalvanic properties of graphene superlattices formed by periodic strain},
	volume = {84},
	url = {https://link.aps.org/doi/10.1103/PhysRevB.84.235440},
	doi = {10.1103/PhysRevB.84.235440},
	number = {23},
	urldate = {2025-07-24},
	journal = {Physical Review B},
	author = {Kiselev, Yu. Yu. and Golub, L. E.},
	month = dec,
	year = {2011},
	pages = {235440},
}

@article{liu_snake_2015,
	title = {Snake states and their symmetries in graphene},
	volume = {92},
	url = {https://link.aps.org/doi/10.1103/PhysRevB.92.235438},
	doi = {10.1103/PhysRevB.92.235438},
	number = {23},
	urldate = {2025-06-03},
	journal = {Physical Review B},
	author = {Liu, Yang and Tiwari, Rakesh P. and Brada, Matej and Bruder, C. and Kusmartsev, F. V. and Mele, E. J.},
	month = dec,
	year = {2015},
	pages = {235438},
}

@article{ahmed_transport_2023,
	title = {Transport in strained graphene: {Interplay} of {Abelian} and axial magnetic fields},
	volume = {108},
	shorttitle = {Transport in strained graphene},
	url = {https://link.aps.org/doi/10.1103/PhysRevB.108.155426},
	doi = {10.1103/PhysRevB.108.155426},
	number = {15},
	urldate = {2025-05-29},
	journal = {Physical Review B},
	author = {Ahmed, Aqeel and Das, Sanjib Kumar and Roy, Bitan},
	month = oct,
	year = {2023},
	pages = {155426},
}

@article{taychatanapat_conductance_2015,
	title = {Conductance oscillations induced by ballistic snake states in a graphene heterojunction},
	volume = {6},
	copyright = {2015 Springer Nature Limited},
	issn = {2041-1723},
	url = {https://www.nature.com/articles/ncomms7093},
	doi = {10.1038/ncomms7093},
	number = {1},
	urldate = {2025-05-28},
	journal = {Nature Communications},
	author = {Taychatanapat, Thiti and Tan, Jun You and Yeo, Yuting and Watanabe, Kenji and Taniguchi, Takashi and \"Ozyilmaz, Barbaros},
	month = feb,
	year = {2015},
	pages = {6093},
}

@article{zwierzycki_calculating_2008,
author = {Zwierzycki, M. and Khomyakov, P. A. and Starikov, A. A. and Xia, K. and Talanana, M. and Xu, P. X. and Karpan, V. M. and Marushchenko, I. and Turek, I. and Bauer, G. E. W. and Brocks, G. and Kelly, P. J.},
title = {Calculating scattering matrices by wave function matching},
journal = {physica status solidi (b)},
volume = {245},
number = {4},
pages = {623-640},
doi = {https://doi.org/10.1002/pssb.200743359},
url = {https://onlinelibrary.wiley.com/doi/abs/10.1002/pssb.200743359},
eprint = {https://onlinelibrary.wiley.com/doi/pdf/10.1002/pssb.200743359},
year = {2008}
}

@article{li_valley_2020,
	title = {Valley {Polarization} and {Inversion} in {Strained} {Graphene} via {Pseudo}-{Landau} {Levels}, {Valley} {Splitting} of {Real} {Landau} {Levels}, and {Confined} {States}},
	volume = {124},
	url = {https://link.aps.org/doi/10.1103/PhysRevLett.124.106802},
	doi = {10.1103/PhysRevLett.124.106802},
	number = {10},
	urldate = {2025-05-21},
	journal = {Physical Review Letters},
	author = {Li, Si-Yu and Su, Ying and Ren, Ya-Ning and He, Lin},
	month = mar,
	year = {2020},
	pages = {106802},
}

@article{de_beule_elastic_2025,
	title = {Elastic {Screening} of {Pseudogauge} {Fields} in {Graphene}},
	volume = {134},
	url = {https://link.aps.org/doi/10.1103/PhysRevLett.134.046404},
	doi = {10.1103/PhysRevLett.134.046404},
	number = {4},
	urldate = {2025-05-20},
	journal = {Physical Review Letters},
	author = {De Beule, Christophe and Smeyers, Robin and Luna, Wilson Nieto and Mele, E. J. and Covaci, Lucian},
	month = jan,
	year = {2025},
	pages = {046404},
}

@article{milovanovic_strain_2019,
	title = {Strain fields in graphene induced by nanopillar mesh},
	volume = {125},
	issn = {0021-8979},
	url = {https://doi.org/10.1063/1.5074182},
	doi = {10.1063/1.5074182},
	number = {8},
	urldate = {2025-05-20},
	journal = {Journal of Applied Physics},
	author = {Milovanovi\'c, S. P. and Covaci, L. and Peeters, F. M.},
	month = jan,
	year = {2019},
	pages = {082534},
}

@article{kapfer_programming_2023,
	title = {Programming twist angle and strain profiles in {2D} materials},
	volume = {381},
	url = {https://www.science.org/doi/10.1126/science.ade9995},
	doi = {10.1126/science.ade9995},
	number = {6658},
	urldate = {2025-05-19},
	journal = {Science},
	author = {Kapfer, Ma\"elle and Jessen, Bjarke S. and Eisele, Megan E. and Fu, Matthew and Danielsen, Dorte R. and Darlington, Thomas P. and Moore, Samuel L. and Finney, Nathan R. and Marchese, Ariane and Hsieh, Valerie and Majchrzak, Paulina and Jiang, Zhihao and Biswas, Deepnarayan and Dudin, Pavel and Avila, Jos\'e and Watanabe, Kenji and Taniguchi, Takashi and Ulstrup, S{\o}ren and B{\o}ggild, Peter and Schuck, P. J. and Basov, Dmitri N. and Hone, James and Dean, Cory R.},
	month = aug,
	year = {2023},
	pages = {677--681},
}

@article{pantaleon_designing_2024,
	title = {Designing {Moir\'e} {Patterns} by {Shearing}},
	volume = {18},
	issn = {1936-0851},
	url = {https://doi.org/10.1021/acsnano.4c08302},
	doi = {10.1021/acsnano.4c08302},
	number = {42},
	urldate = {2025-05-19},
	journal = {ACS Nano},
	author = {Pantale\'on, Pierre A. and Sainz-Cruz, H\'ector and Guinea, Francisco},
	month = oct,
	year = {2024},
	pages = {28575--28584},
}

@article{liu_analytic_2022,
	title = {Analytic solution to pseudo-{Landau} levels in strongly bent graphene nanoribbons},
	volume = {4},
	url = {https://link.aps.org/doi/10.1103/PhysRevResearch.4.023137},
	doi = {10.1103/PhysRevResearch.4.023137},
	number = {2},
	urldate = {2025-05-19},
	journal = {Physical Review Research},
	author = {Liu, Tianyu and Lu, Hai-Zhou},
	month = may,
	year = {2022},
	pages = {023137},
}

@misc{ma_giant_2025,
	title = {Giant elastoresistance in magic-angle twisted bilayer graphene},
	url = {http://arxiv.org/abs/2505.10506},
	doi = {10.48550/arXiv.2505.10506},
	urldate = {2025-05-17},
	publisher = {arXiv},
	author = {Ma, Xuetao and Liu, Zhaoyu and Cai, Jiaqi and Watanabe, Kenji and Taniguchi, Takashi and Xu, Xiaodong and Chu, Jiun-Haw and Yankowitz, Matthew},
	month = may,
	year = {2025},
	note = {arXiv:2505.10506 [cond-mat]},
}

@article{kolasinski_interference_2016,
	title = {Interference features in scanning gate conductance maps of quantum point contacts with disorder},
	volume = {94},
	url = {https://link.aps.org/doi/10.1103/PhysRevB.94.075301},
	doi = {10.1103/PhysRevB.94.075301},
	number = {7},
	urldate = {2025-05-13},
	journal = {Physical Review B},
	author = {Kolasi\'nski, K. and Szafran, B. and Brun, B. and Sellier, H.},
	month = aug,
	year = {2016},
	pages = {075301},
}

@book{datta_electronic_1995,
	address = {Cambridge},
	series = {Cambridge {Studies} in {Semiconductor} {Physics} and {Microelectronic} {Engineering}},
	title = {Electronic {Transport} in {Mesoscopic} {Systems}},
	isbn = {978-0-521-59943-6},
	url = {https://www.cambridge.org/core/books/electronic-transport-in-mesoscopic-systems/1E55DEF5978AA7B843FF70337C220D8B},
	urldate = {2025-05-13},
	publisher = {Cambridge University Press},
	author = {Datta, Supriyo},
	year = {1995},
	doi = {10.1017/CBO9780511805776},
}

@article{ortiz_graphene_2022,
	title = {Graphene nanoelectromechanical systems as valleytronic devices},
	volume = {106},
	url = {https://link.aps.org/doi/10.1103/PhysRevB.106.035416},
	doi = {10.1103/PhysRevB.106.035416},
	number = {3},
	urldate = {2025-05-13},
	journal = {Physical Review B},
	author = {Ortiz, Walter and Szpak, Nikodem and Stegmann, Thomas},
	month = jul,
	year = {2022},
	pages = {035416},
}

@article{chaves_wave-packet_2010,
	title = {Wave-packet dynamics and valley filter in strained graphene},
	volume = {82},
	url = {https://link.aps.org/doi/10.1103/PhysRevB.82.205430},
	doi = {10.1103/PhysRevB.82.205430},
	number = {20},
	urldate = {2025-05-13},
	journal = {Physical Review B},
	author = {Chaves, Andrey and Covaci, L. and Rakhimov, Kh. Yu. and Farias, G. A. and Peeters, F. M.},
	month = nov,
	year = {2010},
	pages = {205430},
}

@article{taychatanapat_electrically_2013,
	title = {Electrically tunable transverse magnetic focusing in graphene},
	volume = {9},
	copyright = {2013 Springer Nature Limited},
	issn = {1745-2481},
	url = {https://www.nature.com/articles/nphys2549},
	doi = {10.1038/nphys2549},
	number = {4},
	urldate = {2025-05-07},
	journal = {Nature Physics},
	author = {Taychatanapat, Thiti and Watanabe, Kenji and Taniguchi, Takashi and Jarillo-Herrero, Pablo},
	month = apr,
	year = {2013},
	pages = {225--229},
}

@article{nigge_room_2019,
	title = {Room temperature strain-induced {Landau} levels in graphene on a wafer-scale platform},
	volume = {5},
	url = {https://www.science.org/doi/10.1126/sciadv.aaw5593},
	doi = {10.1126/sciadv.aaw5593},
	number = {11},
	urldate = {2024-08-12},
	journal = {Science Advances},
	author = {Nigge, P. and Qu, A. C. and Lantagne-Hurtubise, E. and Marsell, E. and Link, S. and Tom, G. and Zonno, M. and Michiardi, M. and Schneider, M. and Zhdanovich, S. and Levy, G. and Starke, U. and Guti\'errez, C. and Bonn, D. and Burke, S. A. and Franz, M. and Damascelli, A.},
	month = nov,
	year = {2019},
	pages = {eaaw5593},
}

@article{rickhaus_snake_2015,
	title = {Snake trajectories in ultraclean graphene p-n junctions},
	volume = {6},
	copyright = {2015 The Author(s)},
	issn = {2041-1723},
	url = {https://www.nature.com/articles/ncomms7470},
	doi = {10.1038/ncomms7470},
	number = {1},
	urldate = {2025-02-16},
	journal = {Nature Communications},
	author = {Rickhaus, Peter and Makk, P\'eter and Liu, Ming-Hao and T\'ov\'ari, Endre and Weiss, Markus and Maurand, Romain and Richter, Klaus and Sch\"onenberger, Christian},
	month = mar,
	year = {2015},
	pages = {6470},
}

@article{neek-amal_electronic_2013,
	title = {Electronic structure of a hexagonal graphene flake subjected to triaxial stress},
	volume = {88},
	url = {https://link.aps.org/doi/10.1103/PhysRevB.88.115428},
	doi = {10.1103/PhysRevB.88.115428},
	number = {11},
	urldate = {2024-01-29},
	journal = {Physical Review B},
	author = {Neek-Amal, M. and Covaci, L. and Shakouri, Kh. and Peeters, F. M.},
	month = sep,
	year = {2013},
	pages = {115428},
}

@article{kang_pseudo-magnetic_2021,
	title = {Pseudo-magnetic field-induced slow carrier dynamics in periodically strained graphene},
	volume = {12},
	copyright = {2021 The Author(s)},
	issn = {2041-1723},
	url = {https://www.nature.com/articles/s41467-021-25304-0},
	doi = {10.1038/s41467-021-25304-0},
	number = {1},
	urldate = {2023-09-20},
	journal = {Nature Communications},
	author = {Kang, Dong-Ho and Sun, Hao and Luo, Manlin and Lu, Kunze and Chen, Melvina and Kim, Youngmin and Jung, Yongduck and Gao, Xuejiao and Parluhutan, Samuel Jior and Ge, Junyu and Koh, See Wee and Giovanni, David and Sum, Tze Chien and Wang, Qi Jie and Li, Hong and Nam, Donguk},
	month = aug,
	year = {2021},
	pages = {5087},
}

@article{guinea_energy_2010,
	title = {Energy gaps and a zero-field quantum {Hall} effect in graphene by strain engineering},
	volume = {6},
	copyright = {2009 Springer Nature Limited},
	issn = {1745-2481},
	url = {https://www.nature.com/articles/nphys1420},
	doi = {10.1038/nphys1420},
	number = {1},
	urldate = {2024-08-12},
	journal = {Nature Physics},
	author = {Guinea, F. and Katsnelson, M. I. and Geim, A. K.},
	month = jan,
	year = {2010},
	pages = {30--33},
}

@article{guinea_generating_2010,
	title = {Generating quantizing pseudomagnetic fields by bending graphene ribbons},
	volume = {81},
	url = {https://link.aps.org/doi/10.1103/PhysRevB.81.035408},
	doi = {10.1103/PhysRevB.81.035408},
	number = {3},
	urldate = {2025-02-08},
	journal = {Physical Review B},
	author = {Guinea, F. and Geim, A. K. and Katsnelson, M. I. and Novoselov, K. S.},
	month = jan,
	year = {2010},
	pages = {035408},
}

@article{cavalcante_all-strain_2016,
	title = {All-strain based valley filter in graphene nanoribbons using snake states},
	volume = {94},
	url = {https://link.aps.org/doi/10.1103/PhysRevB.94.075432},
	doi = {10.1103/PhysRevB.94.075432},
	number = {7},
	urldate = {2025-03-03},
	journal = {Physical Review B},
	author = {Cavalcante, L. S. and Chaves, A. and da Costa, D. R. and Farias, G. A. and Peeters, F. M.},
	month = aug,
	year = {2016},
	pages = {075432},
}

@Article{Jones_Quantized_2017,
author={Jones, Gareth W.
and Bahamon, Dario Andres
and Castro Neto, Antonio H.
and Pereira, Vitor M.},
title={Quantized Transport, Strain-Induced Perfectly Conducting Modes, and Valley Filtering on Shape-Optimized Graphene Corbino Devices},
journal={Nano Letters},
year={2017},
month={Sep},
day={13},
publisher={American Chemical Society},
volume={17},
number={9},
pages={5304-5313},
issn={1530-6984},
doi={10.1021/acs.nanolett.7b01663},
url={https://doi.org/10.1021/acs.nanolett.7b01663}
}

@article{ye_electrons_1995,
	title = {Electrons in a {Periodic} {Magnetic} {Field} {Induced} by a {Regular} {Array} of {Micromagnets}},
	volume = {74},
	url = {https://link.aps.org/doi/10.1103/PhysRevLett.74.3013},
	doi = {10.1103/PhysRevLett.74.3013},
	number = {15},
	urldate = {2025-02-16},
	journal = {Physical Review Letters},
	author = {Ye, P. D. and Weiss, D. and Gerhardts, R. R. and Seeger, M. and von Klitzing, K. and Eberl, K. and Nickel, H.},
	month = apr,
	year = {1995},
	pages = {3013--3016},
}

@article{liu_scalable_2015,
	title = {Scalable {Tight}-{Binding} {Model} for {Graphene}},
	volume = {114},
	url = {https://link.aps.org/doi/10.1103/PhysRevLett.114.036601},
	doi = {10.1103/PhysRevLett.114.036601},
	number = {3},
	urldate = {2025-02-13},
	journal = {Physical Review Letters},
	author = {Liu, Ming-Hao and Rickhaus, Peter and Makk, P\'eter and T\'ov\'ari, Endre and Maurand, Romain and Tkatschenko, Fedor and Weiss, Markus and Sch\"onenberger, Christian and Richter, Klaus},
	month = jan,
	year = {2015},
	pages = {036601},
}

@article{srut_rakic_interaction_2025,
	title = {Interaction {Effects} and {Non}-{Integer} {Pseudo}-{Landau} {Levels} in {Engineered} {Periodically} {Strained} {Graphene}},
	volume = {25},
	issn = {1530-6984},
	url = {https://doi.org/10.1021/acs.nanolett.4c03542},
	doi = {10.1021/acs.nanolett.4c03542},
	number = {1},
	urldate = {2025-02-13},
	journal = {Nano Letters},
	author = {\v{S}rut Raki\'c, Iva and Gilbert, Matthew J. and Sarkar, Preetha and Aishwarya, Anuva and Polini, Marco and Madhavan, Vidya and Mason, Nadya},
	month = jan,
	year = {2025},
	pages = {41--47},
}

@article{milovanovic_strain_2016,
	title = {Strain controlled valley filtering in multi-terminal graphene structures},
	volume = {109},
	issn = {0003-6951},
	url = {https://doi.org/10.1063/1.4967977},
	doi = {10.1063/1.4967977},
	number = {20},
	urldate = {2025-02-13},
	journal = {Applied Physics Letters},
	author = {Milovanovi\'c, S. P. and Peeters, F. M.},
	month = nov,
	year = {2016},
	pages = {203108},
}

@article{pereira_tight-binding_2009,
	title = {Tight-binding approach to uniaxial strain in graphene},
	volume = {80},
	url = {https://link.aps.org/doi/10.1103/PhysRevB.80.045401},
	doi = {10.1103/PhysRevB.80.045401},
	number = {4},
	urldate = {2025-02-08},
	journal = {Physical Review B},
	author = {Pereira, Vitor M. and Castro Neto, A. H. and Peres, N. M. R.},
	month = jul,
	year = {2009},
	pages = {045401},
}

@article{zhang_magnetotransport_2019,
	title = {Magnetotransport in a strain superlattice of graphene},
	volume = {115},
	issn = {0003-6951},
	url = {https://doi.org/10.1063/1.5125462},
	doi = {10.1063/1.5125462},
	number = {14},
	urldate = {2024-12-24},
	journal = {Applied Physics Letters},
	author = {Zhang, Yingjie and Kim, Youngseok and Gilbert, Matthew J. and Mason, Nadya},
	month = oct,
	year = {2019},
	pages = {143508},
}

@article{zhang_gate-tunable_2022,
	title = {Gate-tunable {Veselago} interference in a bipolar graphene microcavity},
	volume = {13},
	copyright = {2022 The Author(s)},
	issn = {2041-1723},
	url = {https://www.nature.com/articles/s41467-022-34347-w},
	doi = {10.1038/s41467-022-34347-w},
	number = {1},
	urldate = {2025-02-03},
	journal = {Nature Communications},
	author = {Zhang, Xi and Ren, Wei and Bell, Elliot and Zhu, Ziyan and Tsai, Kan-Ting and Luo, Yujie and Watanabe, Kenji and Taniguchi, Takashi and Kaxiras, Efthimios and Luskin, Mitchell and Wang, Ke},
	month = nov,
	year = {2022},
	pages = {6711},
}

@article{nguyen_valley_2016,
  title = {{Valley Filtering and Electronic Optics Using Polycrystalline Graphene}},
  author = {Nguyen, V. Hung and Dechamps, S. and Dollfus, P. and Charlier, J.-C.},
  journal = {Phys. Rev. Lett.},
  volume = {117},
  issue = {24},
  pages = {247702},
  numpages = {6},
  year = {2016},
  month = {Dec},
  publisher = {American Physical Society},
  doi = {10.1103/PhysRevLett.117.247702},
  url = {https://link.aps.org/doi/10.1103/PhysRevLett.117.247702}
}

@article{zhu_programmable_2015,
	title = {{Programmable Extreme Pseudomagnetic Fields in Graphene by a Uniaxial Stretch}},
	volume = {115},
	url = {https://link.aps.org/doi/10.1103/PhysRevLett.115.245501},
	doi = {10.1103/PhysRevLett.115.245501},
	number = {24},
	urldate = {2024-08-12},
	journal = {Physical Review Letters},
	author = {Zhu, Shuze and Stroscio, Joseph A. and Li, Teng},
	month = dec,
	year = {2015},
	pages = {245501},
}

@article{zhang_electronic_2018,
	title = {Electronic transport in a two-dimensional superlattice engineered via self-assembled nanostructures},
	volume = {2},
	copyright = {2018 The Author(s)},
	issn = {2397-7132},
	url = {https://www.nature.com/articles/s41699-018-0076-0},
	doi = {10.1038/s41699-018-0076-0},
	number = {1},
	urldate = {2024-01-29},
	journal = {npj 2D Materials and Applications},
	author = {Zhang, Yingjie and Kim, Youngseok and Gilbert, Matthew J. and Mason, Nadya},
	month = sep,
	year = {2018},
	pages = {1--6},
}

@article{woods_electron-phonon_2000,
	title = {Electron-phonon effects in graphene and armchair (10,10) single-wall carbon nanotubes},
	volume = {61},
	url = {https://link.aps.org/doi/10.1103/PhysRevB.61.10651},
	doi = {10.1103/PhysRevB.61.10651},
	number = {16},
	urldate = {2024-07-17},
	journal = {Physical Review B},
	author = {Woods, L. M. and Mahan, G. D.},
	month = apr,
	year = {2000},
	pages = {10651--10663},
}

@article{vozmediano_gauge_2010,
	title = {Gauge fields in graphene},
	volume = {496},
	issn = {0370-1573},
	url = {https://www.sciencedirect.com/science/article/pii/S0370157310001729},
	doi = {10.1016/j.physrep.2010.07.003},
	number = {4},
	urldate = {2023-09-24},
	journal = {Physics Reports},
	author = {Vozmediano, M. A. H. and Katsnelson, M. I. and Guinea, F.},
	month = nov,
	year = {2010},
	pages = {109--148},
}

@article{suzuura_phonons_2002,
	title = {Phonons and electron-phonon scattering in carbon nanotubes},
	volume = {65},
	url = {https://link.aps.org/doi/10.1103/PhysRevB.65.235412},
	doi = {10.1103/PhysRevB.65.235412},
	number = {23},
	urldate = {2023-09-24},
	journal = {Physical Review B},
	author = {Suzuura, Hidekatsu and Ando, Tsuneya},
	month = may,
	year = {2002},
	pages = {235412},
}

@article{settnes_pseudomagnetic_2016,
	title = {Pseudomagnetic fields and triaxial strain in graphene},
	volume = {93},
	url = {https://link.aps.org/doi/10.1103/PhysRevB.93.035456},
	doi = {10.1103/PhysRevB.93.035456},
	number = {3},
	urldate = {2024-02-15},
	journal = {Physical Review B},
	author = {Settnes, Mikkel and Power, Stephen R. and Jauho, Antti-Pekka},
	month = jan,
	year = {2016},
	pages = {035456},
}

@article{milovanovic_band_2020,
	title = {Band flattening in buckled monolayer graphene},
	volume = {102},
	url = {https://link.aps.org/doi/10.1103/PhysRevB.102.245427},
	doi = {10.1103/PhysRevB.102.245427},
	number = {24},
	urldate = {2024-07-31},
	journal = {Physical Review B},
	author = {Milovanovi\'c, S. P. and An{\dj}elkovi\'c, M. and Covaci, L. and Peeters, F. M.},
	month = dec,
	year = {2020},
	pages = {245427},
}

@article{mao_evidence_2020,
	title = {Evidence of flat bands and correlated states in buckled graphene superlattices},
	volume = {584},
	copyright = {2020 The Author(s), under exclusive licence to Springer Nature Limited},
	issn = {1476-4687},
	url = {https://www.nature.com/articles/s41586-020-2567-3},
	doi = {10.1038/s41586-020-2567-3},
	number = {7820},
	urldate = {2023-09-24},
	journal = {Nature},
	author = {Mao, Jinhai and Milovanovi\'c, Slavi\v{s}a P. and An{\dj}elkovi\'c, Mi\v{s}a and Lai, Xinyuan and Cao, Yang and Watanabe, Kenji and Taniguchi, Takashi and Covaci, Lucian and Peeters, Francois M. and Geim, Andre K. and Jiang, Yuhang and Andrei, Eva Y.},
	month = aug,
	year = {2020},
	pages = {215--220},
}

@article{manesco_correlation-induced_2021,
	title = {Correlation-induced valley topology in buckled graphene superlattices},
	volume = {8},
	issn = {2053-1583},
	url = {https://dx.doi.org/10.1088/2053-1583/ac0b48},
	doi = {10.1088/2053-1583/ac0b48},
	number = {3},
	urldate = {2023-09-20},
	journal = {2D Materials},
	author = {Manesco, Antonio L. R. and Lado, Jose L.},
	month = jun,
	year = {2021},
	pages = {035057},
}

@article{manesco_correlations_2020,
	title = {Correlations in the elastic {Landau} level of spontaneously buckled graphene},
	volume = {8},
	issn = {2053-1583},
	url = {https://dx.doi.org/10.1088/2053-1583/abbc5f},
	doi = {10.1088/2053-1583/abbc5f},
	number = {1},
	urldate = {2023-09-20},
	journal = {2D Materials},
	author = {Manesco, A. L. R. and Lado, J. L. and Ribeiro, E. V. S. and Weber, G. and Jr, D. Rodrigues},
	month = oct,
	year = {2020},
	pages = {015011},
}

@article{levy_strain-induced_2010,
	title = {Strain-{Induced} {Pseudo}-{Magnetic} {Fields} {Greater} {Than} 300 {Tesla} in {Graphene} {Nanobubbles}},
	volume = {329},
	url = {https://www.science.org/doi/10.1126/science.1191700},
	doi = {10.1126/science.1191700},
	number = {5991},
	urldate = {2024-08-12},
	journal = {Science},
	author = {Levy, N. and Burke, S. A. and Meaker, K. L. and Panlasigui, M. and Zettl, A. and Guinea, F. and Neto, A. H. Castro and Crommie, M. F.},
	month = jul,
	year = {2010},
	pages = {544--547},
}

@article{ho_hall_2021,
	title = {Hall effects in artificially corrugated bilayer graphene without breaking time-reversal symmetry},
	volume = {4},
	copyright = {2021 The Author(s), under exclusive licence to Springer Nature Limited},
	issn = {2520-1131},
	url = {https://www.nature.com/articles/s41928-021-00537-5},
	doi = {10.1038/s41928-021-00537-5},
	number = {2},
	urldate = {2023-09-28},
	journal = {Nature Electronics},
	author = {Ho, Sheng-Chin and Chang, Ching-Hao and Hsieh, Yu-Chiang and Lo, Shun-Tsung and Huang, Botsz and Vu, Thi-Hai-Yen and Ortix, Carmine and Chen, Tse-Ming},
	month = feb,
	year = {2021},
	pages = {116--125},
}

@article{gao_untwisting_2023,
	title = {Untwisting {Moir\'e} {Physics}: {Almost} {Ideal} {Bands} and {Fractional} {Chern} {Insulators} in {Periodically} {Strained} {Monolayer} {Graphene}},
	volume = {131},
	shorttitle = {Untwisting {Moir}{\textbackslash}'e {Physics}},
	url = {https://link.aps.org/doi/10.1103/PhysRevLett.131.096401},
	doi = {10.1103/PhysRevLett.131.096401},
	number = {9},
	urldate = {2023-09-20},
	journal = {Physical Review Letters},
	author = {Gao, Qiang and Dong, Junkai and Ledwith, Patrick and Parker, Daniel and Khalaf, Eslam},
	month = aug,
	year = {2023},
	pages = {096401},
}

@article{de_beule_roses_2023,
	title = {Roses in the nonperturbative current response of artificial crystals},
	volume = {120},
	url = {https://www.pnas.org/doi/10.1073/pnas.2306384120},
	doi = {10.1073/pnas.2306384120},
	number = {43},
	urldate = {2024-01-24},
	journal = {Proceedings of the National Academy of Sciences},
	author = {De Beule, Christophe and Phong, V\~o Ti\'\ecircumflex{}n and Mele, E. J.},
	month = oct,
	year = {2023},
	pages = {e2306384120},
}

@article{de_beule_network_2023,
	title = {Network model for periodically strained graphene},
	volume = {107},
	url = {https://link.aps.org/doi/10.1103/PhysRevB.107.045405},
	doi = {10.1103/PhysRevB.107.045405},
	number = {4},
	urldate = {2023-09-24},
	journal = {Physical Review B},
	author = {De Beule, Christophe and Phong, V\~o Ti\'\ecircumflex{}n and Mele, E. J.},
	month = jan,
	year = {2023},
	pages = {045405},
}

@Misc{Note1,
  note = {See Supplemental Material at [insert url] for two-terminal conductance in bent graphene ribbon, and the TPMF in triaxially strained graphene.}
}

@Article{liu_2026_scalable,
 title = {Scalable tight-binding model for strained graphene},
  author = {Liu, Ming-Hao and De Beule, Christophe and Mre\ifmmode \acute{n}\else \'{n}\fi{}ca-Kolasi\ifmmode \acute{n}\else \'{n}\fi{}ska, Alina and Wu, Hsin-You and Garcia-Ruiz, Aitor and Kochan, Denis and Richter, Klaus},
  journal = {Phys. Rev. B},
  volume = {113},
  issue = {19},
  pages = {195429},
  numpages = {13},
  year = {2026},
  month = {May},
  publisher = {American Physical Society},
  doi = {10.1103/ntf2-5dyw},
  url = {https://link.aps.org/doi/10.1103/ntf2-5dyw}
}

@Article{Zhou2025,
author={Zhou, Bowen
and Mre{\'{n}}ca-Kolasi{\'{n}}ska, Alina
and Watanabe, Kenji
and Taniguchi, Takashi},
title={Emergent Thermal Strain-Induced Pseudomagnetic Fields and Shubnikov--de Haas Beating Patterns in Encapsulated Graphene in Extraordinary Magnetoresistance Geometry},
journal={ACS Nano},
year={2025},
month={Aug},
day={19},
publisher={American Chemical Society},
volume={19},
number={32},
pages={29276-29285},
issn={1936-0851},
doi={10.1021/acsnano.5c04844},
url={https://doi.org/10.1021/acsnano.5c04844}
}

@Article{Bhandari2016,
author={Bhandari, Sagar
and Lee, Gil-Ho
and Klales, Anna
and Watanabe, Kenji
and Taniguchi, Takashi
and Heller, Eric
and Kim, Philip
and Westervelt, Robert M.},
title={Imaging Cyclotron Orbits of Electrons in Graphene},
journal={Nano Letters},
year={2016},
month={Mar},
day={09},
publisher={American Chemical Society},
volume={16},
number={3},
pages={1690-1694},
issn={1530-6984},
doi={10.1021/acs.nanolett.5b04609},
url={https://doi.org/10.1021/acs.nanolett.5b04609}
}

@article{beenakker_mode_1988,
doi = {10.1209/0295-5075/7/4/013},
url = {https://doi.org/10.1209/0295-5075/7/4/013},
year = {1988},
month = {oct},
publisher = {},
volume = {7},
number = {4},
pages = {359},
author = {C. W. J. Beenakker and H. van Houten and B. J. van Wees},
title = {Mode Interference Effect in Coherent Electron Focusing},
journal = {Europhysics Letters}
}

@article{van-Houten_coherent_1989,
  title = {Coherent electron focusing with quantum point contacts in a two-dimensional electron gas},
  author = {van Houten, H. and Beenakker, C. W. J. and Williamson, J. G. and Broekaart, M. E. I. and van Loosdrecht, P. H. M. and van Wees, B. J. and Mooij, J. E. and Foxon, C. T. and Harris, J. J.},
  journal = {Phys. Rev. B},
  volume = {39},
  issue = {12},
  pages = {8556--8575},
  numpages = {0},
  year = {1989},
  month = {Apr},
  publisher = {American Physical Society},
  doi = {10.1103/PhysRevB.39.8556},
  url = {https://link.aps.org/doi/10.1103/PhysRevB.39.8556}
}



\clearpage
\onecolumngrid
\begin{center}
\textbf{\Large Supporting Information for ``Pseudomagnetic focusing of valley-polarized beams in strained graphene''}
\end{center}
 
\setcounter{equation}{0}
\setcounter{figure}{0}
\setcounter{table}{0}
\setcounter{page}{1}
\setcounter{secnumdepth}{2}
\makeatletter
\renewcommand{\thepage}{S\arabic{page}}
\renewcommand{\thesection}{S\arabic{section}}
\renewcommand{\theequation}{S\arabic{equation}}
\renewcommand{\thefigure}{S\arabic{figure}}
\renewcommand{\thetable}{S\arabic{table}}

\tableofcontents
\vspace{1cm}

\twocolumngrid


\section{Two-terminal conductance}

\begin{figure}[t]
    \centering
    \includegraphics[width=\columnwidth]{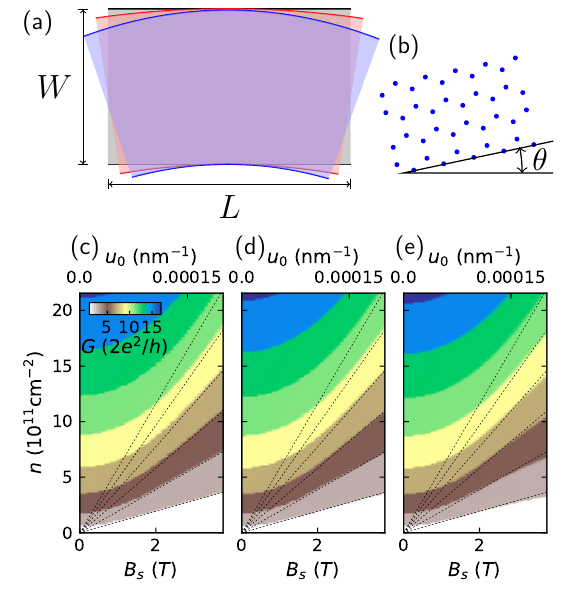}
    \caption{(a) Schematic deformed unscaled graphene lattice with the scaling factor $s=1$ (red) and $s=2$ (blue) with the deformation upscaled according to the scaling method. (b) Definition of the angle $\theta$; Throughout the paper we use $\theta=0$. (c)--(e) Two-terminal conductance of the bent zigzag nanoribbon as a function of the strain and carrier density with (c) $s=1$, (d) $s=2$, and (e) $s=4$. The lower $x$-axis is the strain converted to pseudomagnetic field.}
    \label{figS  two terminal conductance}
\end{figure}

To demonstrate the performance of the scalable tight binding model for the pseudomagnetic field description, we consider a zigag nanoribbon with the deformation profile \cite{guinea_generating_2010}
\begin{equation}
    \begin{aligned}
        u_x &= u_0 \left( 2y + \frac{f_0}{f_1} \right) x, \\
        u_y &= u_0 \left[ \frac{\nu}{\nu - 2} \left( y + \frac{f_0}{f_1} \right) y - x^2 \right],
    \end{aligned}
    \label{eq ur uy for bent graphene}
\end{equation}
where $f_0$ and $f_1$ are constants describing forces acting at the right and left boundaries, we take $\nu = 0.15$ for the Poisson ratio of graphene, and $u_0$ controls the distortion. Here, we used that for a two-dimensional elastic membrane,
\begin{equation}
    \mu = \frac{E}{2 ( 1 + \nu )}, \quad \lambda = \frac{\nu E}{1 - \nu^2}, \quad \nu = \frac{1}{1 + 2 \mu / \lambda},
\end{equation}
with $E$ the Young modulus. The resulting pseudomagnetic field is spatially uniform, which can be seen by calculating the strain tensor
\begin{equation}
    \begin{aligned}
        \epsilon_{xx} &= u_0 \frac{\partial }{\partial x} \left (2xy + \frac{f_0}{f_1}x \right) = u_0 \left ( 2y + \frac{f_0}{f_1} \right), \\
        \epsilon_{yy} &= u_0 \frac{\partial }{\partial y} \left [ \frac{\nu}{\nu - 2} \left( y^2 + \frac{f_0}{f_1}y \right) -x^2 \right] \\
        &= - u_0 \frac{\nu}{2 - \nu} \left( 2y + \frac{f_0}{f_1}
         \right) ,
         \\
         \epsilon_{xy} &= \frac{1}{2}  \left\{
        u_0 \frac{\partial }{\partial y} \left (2xy + \frac{f_0}{f_1}x \right) \right.\\
        &+\left. u_0 \frac{\partial }{\partial x} \left [ \frac{\nu}{\nu - 2} \left( y^2 + \frac{f_0}{f_1}y \right) - x^2 \right]
         \right\} \\
         &= \frac{u_0 }{2} \left(  2x - 2x \right) = 0,
        \end{aligned}
\end{equation}
and from this we obtain the pseudo vector potential (in the nearest-neighbor approximation)
\begin{align}
    \mathbf{A}_s & = -\frac{\hbar\beta}{2ea_0}
    R(3\theta) \begin{pmatrix}
        \epsilon_{xx}-\epsilon_{yy} \\ -2\epsilon_{xy}
    \end{pmatrix} \\
    & = -\frac{\hbar\beta u_0}{ea_0(2 - \nu)}
    R(3\theta) \begin{pmatrix}
        2 y + f_0/f_1 \\ 0
    \end{pmatrix}\ ,
\end{align}
with $a_0 \approx 0.142$~nm the nearest-neighbor distance of pristine graphene and $\theta$ is the angle between the zigzag direction and the $x$ axis [see Fig.\ \ref{figS  two terminal conductance}(b)]. Finally, the pseudo magnetic field becomes
\begin{equation}
    B_s = \frac{\partial A_y}{\partial x}-\frac{\partial A_x}{\partial y} 
    = \frac{2\hbar\beta}{ea_0} \frac{u_0}{2 - \nu} \cos(3\theta) \ .
    \label{eq Bs of an arch}
\end{equation}

The deformed nanoribbon based on this prescription is illustrated in Fig.\ \ref{figS  two terminal conductance}(a). The dimensions marked in Fig.\ \ref{figS  two terminal conductance}(a) are $W\approx 53$ nm and $L\approx 190$ nm. In Fig.\ \ref{figS  two terminal conductance}(a) we show the comparison of lattices with the scaling factor $s=1$ and $s=2$ and the deformation resulting from the scaling approach necessary to keep the pseudomagnetic field unchanged in the scaled system. We assume that the left and right edges of the ribbon are connected to semi-infinite leads, and calculate the two-terminal conductance between them as a function of $u_0$ and carrier density. The results are shown in  Figs.\ \ref{figS  two terminal conductance}(c)--(e) for $s=1, 2, 4$. The maps show regions of constant conductance \cite{ahmed_transport_2023}, quantized at integer values following the sequence expected for quantum Hall plateaus in graphene, except that here no external magnetic field is applied. The $u_0$ values can be converted to a pseudomagnetic field using Eq.\ \eqref{eq Bs of an arch}.
For plotting the pseudo Landau levels (pLL), $B_s$ is corrected by a factor 1.1, due to possible finite size effects on the positions of the conductance steps. The same pLL are displayed in the scans at $s=2$ and $s=4$ [see dashed lines in Figs.\ \ref{figS  two terminal conductance}(c)--(e)]. With $s=2$ [Fig.\ \ref{figS  two terminal conductance}(d)], the steps are at nearly the same positions compared to the unscaled system [Fig.\ \ref{figS  two terminal conductance}(c)], however, when $s=4$ [Fig.\ \ref{figS  two terminal conductance}(e)], the plateaus start to deviate from the expected relation at $B_s \approx 3$ T. This qualitatively confirms that simulations up to that $B_s$ value are reliable at least with $s\leq 4$.
\begin{figure}[t]
    \centering
    \includegraphics[width=\columnwidth]{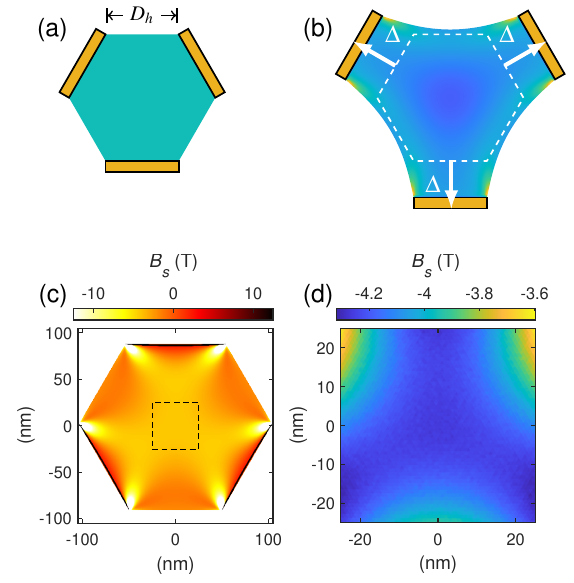}
    \caption{Schematics of a hexagonally shaped graphene flake of edge $D_h$ (a) before  and (b) after the triaxial strain is applied. (b) An exemplary pseudomagnetic field profile considering $(D_h,\Delta)=(100,5)\unit{nm}$, where the central region marked by a black dashed box is magnified in (d).}
    \label{figS triaxial schematic}
\end{figure}
\begin{figure*}
    \centering
    \includegraphics[scale=1]{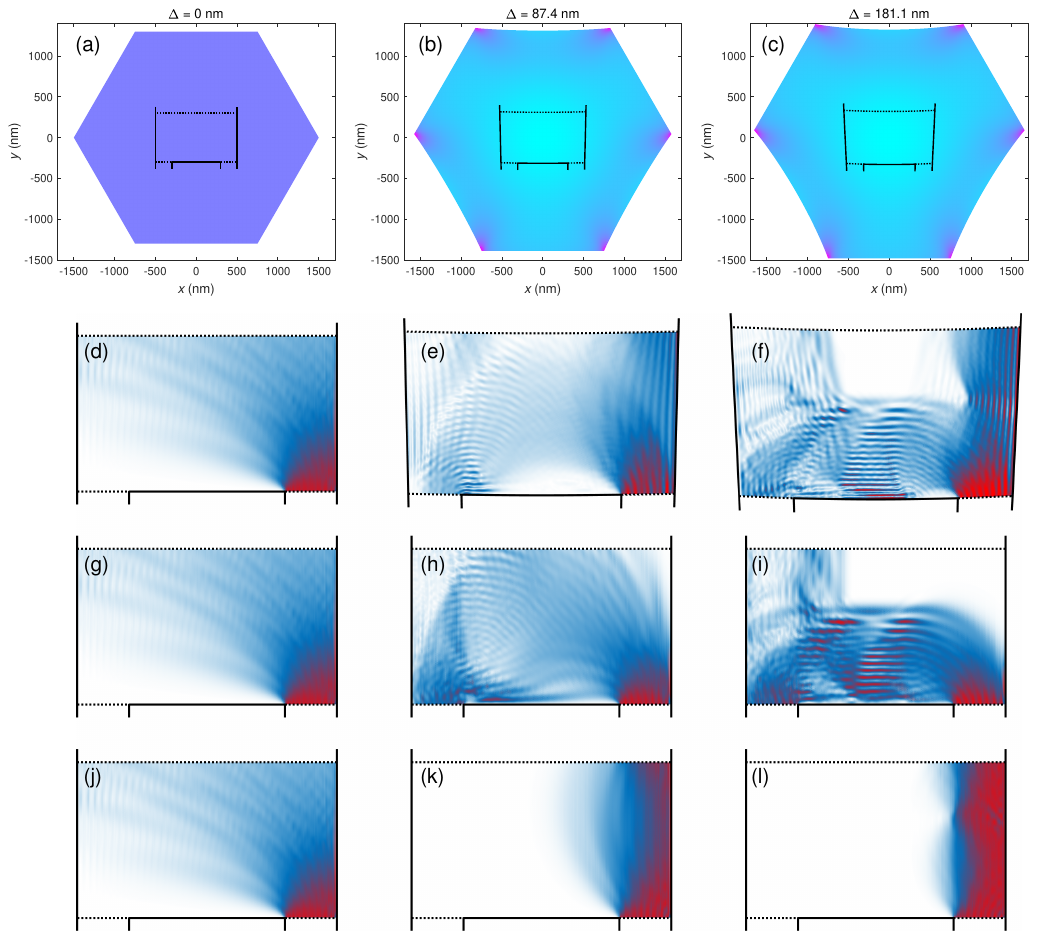}
    \caption{A hexagonally shaped graphene lattice with $D_h=1500\unit{nm}$ under (a) zero strain and (b)--(c) triaxial strain with the values of $\Delta$ shown in the title of the panels; here, $D_h$ and $\Delta$ are defined in Fig.\ \ref{figS triaxial schematic} (a-b). (d)--(f) Current density profiles showing transverse pseudomagnetic focusing. The values of $\Delta$ in (d), (e), and (f) correspond to those in (a), (b), and (c), respectively. (g)--(i) Current density profiles in an unstrained graphene with the geometry sketched at the center of (a) under $B_z=0,0.2\unit{T},0.4\unit{T}$, respectively. The Fermi energy in (d)--(i) is $E=80\unit{meV}$. (j)--(l) are the same as (g)--(i) except $E=-80\unit{meV}$.
    The scaling factor is $s=6$ in all of (d)--(l), sharing the same color tone that white and red means zero and high local current density, respectively.}
    \label{figS Jia-Tong TMF}
\end{figure*}
\begin{figure*}
    \centering
    \includegraphics[width=\textwidth]{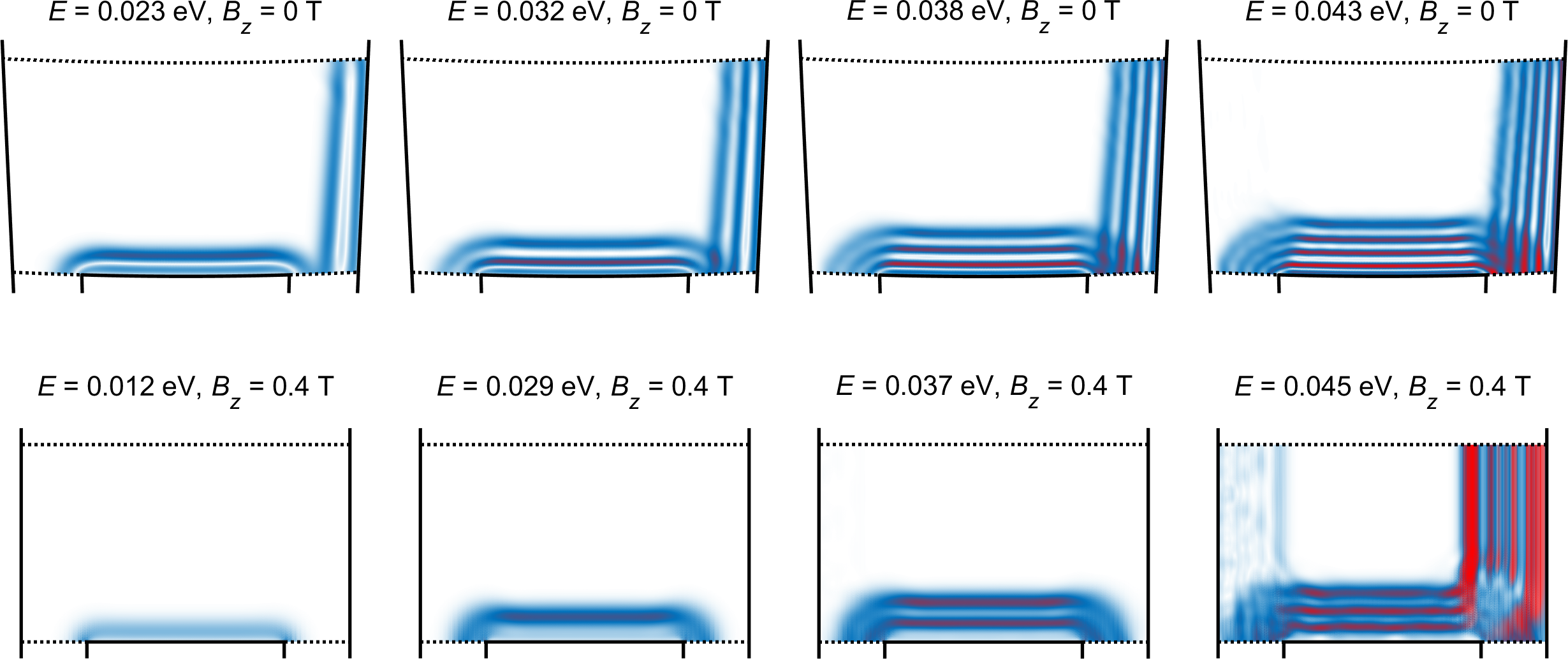}
    \caption{Upper row considers the same parameters as Fig.\ \ref{figS Jia-Tong TMF}(f) except for lower $E$ shown in the title of each panel. Lower row considers the same parameters as Fig.\ \ref{figS Jia-Tong TMF}(i) except for lower $E$ shown in the title of each panel. The scaling factor for all panels is $s=6$.}
    \label{figS Jia-Tong QH}
\end{figure*}



\section{Transport under a triaxial strain field}

Here, we consider triaxially-strained graphene to show another series of test calculations to reveal pseudomagnetotransport. These were obtained by computing the transmission function with the real-space Green's function method, and the Keldysh-Green's function method in the linear response regime for the calculation of local current densities \cite{datta_electronic_1995}. 
The schematic of the geometry considered is shown in Figs.\ \ref{figS triaxial schematic}(a)--(b), where a hexagonal flake of side length $D_h$ and zigzag horizontal edge is deformed by displacing three of the sides by $\Delta$ in the normal direction \cite{guinea_energy_2010}. Exemplary PMF spatial profile maps resulting in the flake is shown in Fig.\ \ref{figS triaxial schematic}(c) and a zoom in of the central region marked with a box therein is shown in Fig.\ \ref{figS triaxial schematic}(d). As seen in Fig.\ \ref{figS triaxial schematic}(d), the PMF varies only slightly in the central region.
To study transport in such weakly spatially varying PMF, we start with a hexagon-shaped graphene flake, as shown in Fig.\ \ref{figS Jia-Tong TMF}(a), and focus on the central region marked by the outline of a fictitious 3-lead device. When the hexagonal flake undergoes a triaxial strain, such as Figs.\ \ref{figS Jia-Tong TMF}(b) and (c), the fictitious device outline, and hence the lattice sites therein, deform correspondingly. We then inject current from the bottom right lead and image the local current densities. Images of current density profiles corresponding to Figs.\ \ref{figS Jia-Tong TMF}(a)--(c) without applied $B_z$ are shown in Figs.\ \ref{figS Jia-Tong TMF}(d)--(f), respectively, considering Fermi energy $E=80\unit{meV}$. As seen in panels (e) and (f) of Fig.\ \ref{figS Jia-Tong TMF}, transverse pseudomagnetic focusing can be clearly seen. Moreover, in addition to the focused current that is bent to the target of the bottom left lead, another fraction of the injected current bent to the opposite direction can also be seen, due to the opposite PMF sensed by the other valley.

Figures \ref{figS Jia-Tong TMF}(g)--(i) consider no strain but an applied magnetic field: $B_z=0$ in (g), $B_z=0.2\unit{T}$ in (h), and $B_z=0.4\unit{T}$ in (i). Note that Fig.\ \ref{figS Jia-Tong TMF}(d) and Fig.\ \ref{figS Jia-Tong TMF}(g) are simply a sanity check because they were computed using different codes. In contrast to Figs.\ \ref{figS Jia-Tong TMF}(e) and (f) where the injected current bends toward both left and right, panels (h) and (i) show only the left bending current consistent with the direction of the Lorentz force. When the sign of the carriers is changed by considering $E=-80\unit{meV}$, the left-bending current becomes right-bending, as shown in Figs.\ \ref{figS Jia-Tong TMF}(j)--(l). Again, panels (g) and (j) are for a sanity check, revealing identical current density profiles for $E=\pm 80\unit{meV}$ at $B_z=0$, due to the electron-hole symmetry present in graphene with only nearest neighbor hopping.

Next, we focus on Fig.\ \ref{figS Jia-Tong TMF}(i) and lower the energy so that the system enters the quantum Hall regime, as shown in lower row of the panels in Fig.\ \ref{figS Jia-Tong QH}, where standard quantum Hall edge currents guided toward the bottom left lead can be seen. In contrast, if we focus on Fig.\ \ref{figS Jia-Tong TMF}(f) for a strained graphene free of external magnetic field and consider lower energies, half of the pseudo-quantum Hall edge current shown in the upper panels of Fig.\ \ref{figS Jia-Tong QH} is guided to the bottom left lead, and the other half travels along the right edge. These oppositely propagating pseudo-quantum Hall edge currents are expected to be fully valley-polarized.

Although tailoring a rectangular sample from a triaxially strained hexagon-shaped graphene sample is rather unrealistic, these simulations summarized in Fig.\ \ref{figS Jia-Tong TMF} and Fig.\ \ref{figS Jia-Tong QH} perfectly reveal large-scale pseudomagnetotransport including transverse pseudomagnetic focusing and the pseudo-quantum Hall effect.

\section{Estimation of strain in the bent geometry}

\begin{figure*}[t]
    \centering
    \includegraphics[width=\textwidth]{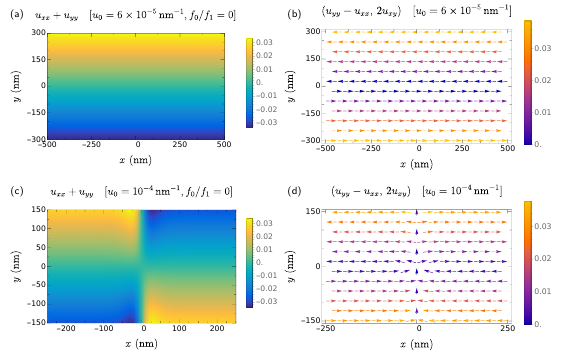}
    \caption{(a) Trace of the strain tensor $u_{xx}+u_{yy}$ and (b) shear strains for the bent-ribbon geometry taking the largest value $u_0=6\times 10^{-5}$ nm$^{-1}$ considered in the main text. (c)--(d) same as (a)--(b) in the S-shaped ribbon for $u_0=1\times 10^{-4}$ nm$^{-1}$. Here, only the trace depends on the ratio $f_0/f_1$ which is related to boundary conditions \cite{guinea_generating_2010} and which is set to zero for concreteness.}
    \label{figS strain}
\end{figure*}

In order to determine the feasibility of the proposed geometry, we estimate the strain magnitude required to observe conductance oscillations in PTMF and the S-bent ribbon. We visualize the strain in Fig.\ \ref{figS strain}. In particular, we show both the trace of the strain tensor, $u_{xx}+u_{yy}$, which corresponds to local expansion and compression, and the shear strains $(u_{yy} - u_{xx}, 2 u_{xy})$. These are shown in Figs.\ \ref{figS strain}(a) and (b) for the bent-ribbon geometry taking the largest displacement magnitude $u_0=6\times 10^{-5}$ nm$^{-1}$ used in the simulation. The strain varies between $\pm 3\%$, which is well within the elastic regime for graphene. Similarly, Figs.\ \ref{figS strain}(c) and (d) present a similar check at $u_0=1\times 10^{-4}$ nm$^{-1}$ for the S-bent ribbon, displaying again a maximum strain of 3\%. Here, we take the value of $u_0$ relevant to our convergence checks for different scaling parameters in Section \ref{sectionS4}.

\section{Role of edge roughness and scaling}
\label{sectionS4}

\begin{figure}[t]
    \centering
    \includegraphics[width=\columnwidth]{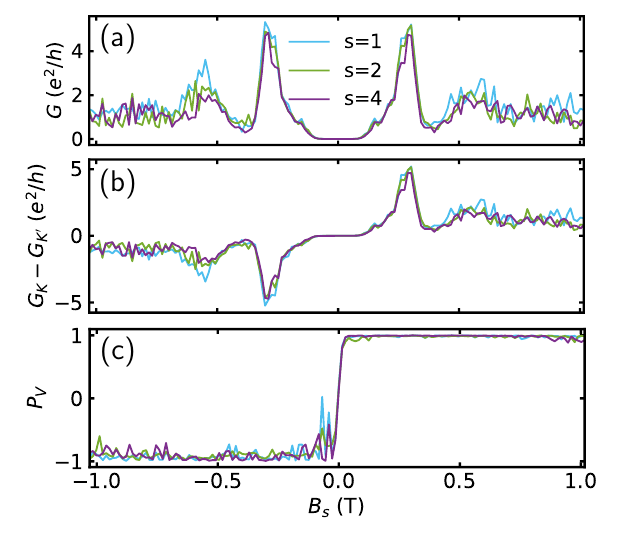}
    \caption{(a) Injector to collector conductance as a function of $B_s$ with edge disorder and various $s$. (b) $G_{\mathrm{K}}-G_{\mathrm{K'}}$  and (c) $P_V$ in same system as (a).}
    \label{figS edge}
\end{figure}

 Edge roughness is highly relevant to our proposed device due to its ability to induce intervalley scattering. To address the issue of the edge disorder, we performed several simulations in the armchair geometry shown in Fig.\ 2(e) of the main text, where atomic sites near the right boundary of the system were randomly removed with a probability of 0.4. We compared the effect of such roughness for the scaling factors $s=1,2,4$. The results are shown in Fig.\ \ref{figS edge}. As a consequence of the edge disorder, the $\mathbf{K}$ ($\mathbf{K'}$) beam hitting the edge between the leads is not completely reflected to $\mathbf{K'}$ ($\mathbf{K}$) valley. Instead, 
 scattering from the edge partially mixes valleys, so there is nonzero conductance beyond the first peak. 
 The second focusing peak becomes visible, albeit somewhat broadened due to the loss of specular reflection caused by the edge roughness. 

Note that edge roughness on the lower (zigzag) edge can be detrimental to valley polarization since the beams reflected from $\mathbf{K}$ to $\mathbf{K'}$ or vice versa may reach the collector lead. This can be avoided by placing a current drain at the lower side of the sample.

Here, we can also comment on the role of the scaling factor $s$. One must note that the effect of removing edge atoms varies in the scaled and unscaled lattices, therefore specific fluctuations vary for each $s$.
Nevertheless, the scaling factor does not affect the overall high valley polarization. This is because the pseudomagnetic field $B_s$ steers the electrons belonging to the two valleys in the opposite directions.

\begin{figure}[t]
    \centering
    \includegraphics[width=\columnwidth]{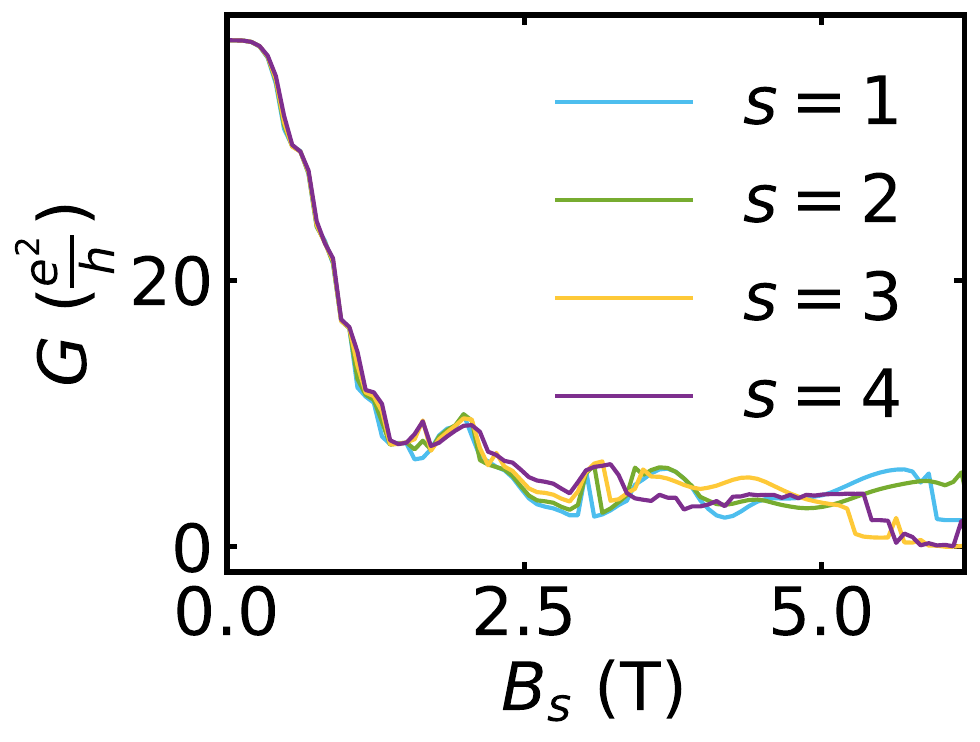}
    \caption{2-terminal conductance in the S-bent ribbon at $E=0.1$ eV as a function of $B_s$ with various $s$.}
    \label{figS ssnake}
\end{figure}

 In addition to the scaling factor analysis in the presence of edge roughness, we performed rigorous tests in the S-shaped ribbon. 
Figure \ref{figS ssnake} shows the conductance line cut at $E=0.1$ eV with $s=1,2,3,4$. Crucially, the results with $s=2$ (used in Fig.\ 4 of the main text) are in good agreement with the baseline $s=1$, confirming that the observed phenomena--including peak positions and heights--are not artifacts of the scaling procedure. At higher $s$, the curves begin to deviate from baseline at a threshold $B_s$. This threshold field decreases as the scaling factor $s$ increases. For instance, the trace $s=3$ maintains a good agreement with the baseline up to $u_0\approx 1\times 10^{-4}$ nm, corresponding to a maximum strain of 3\% [see Fig.\ \ref{figS strain}. This corresponds to $s u_{ij}\approx 0.09$, in accordance with the estimations of \cite{liu_2026_scalable}.
 
\section{Revisiting the TMF experiment in graphene}
\begin{figure}[t]
    \centering
    \includegraphics[width=\columnwidth]{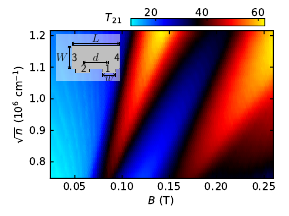}
    \caption{Calculated injector to collector transmission as a function of magnetic field $B$ and square root of carrier density $n^{1/2}$ revisiting the TMF experiment on graphene \cite{Bhandari2016}. The inset shows the schematic of the system used for the simulation. }
    \label{figS tmf}
\end{figure}

To show the reliability of our quantum transport simulations for realistic devices, we performed the quantum transport calculation of the transverse magnetic focusing spectrum corresponding to the multi-terminal system investigated in \cite{Bhandari2016}. Following the dimensions of the experimental device, we used a system shown in the inset of Fig.\ \ref{figS tmf}, with $d=2\ \mu$m, $w=0.7\ \mu$m, $W=1.75\ \mu$m, and $L=3.06\ \mu$m, using $s=10$. The conductance peaks positions are in good agreement with the experimental counterpart in Fig.\ 2(a) of \cite{Bhandari2016}. Note that the overall magnitude can differ between the injector to collector conductance in our map, and the resistance variation in the measurement. By computing all transmission functions to get the four-terminal nonlocal resistance the simulation can become quantitatively comparable with the experiment. For a comparison of numerical transmission function and 4-point resistance, see Methods section of \cite{Mrenca2023_probing}, in which we recover the first TMF experiment \cite{taychatanapat_electrically_2013}.
These transport simulations validate our transport methodology as well as the single-electron approximation used throughout the paper.

\end{document}